# Does Private Equity Hurt or Improve Healthcare Value? New Evidence and Mechanisms

**Minghong Yuan**
University of Texas at Austin
2110 Speedway
Austin, Texas
United States
78712
minghong.yuan@mccombs.utexas.edu

**Wen Wen**
University of Texas at Austin
2110 Speedway
Austin, Texas
United States
78712
Wen.Wen@mccombs.utexas.edu

**Indranil Bardhan**
University of Texas at Austin
2110 Speedway
Austin, Texas
United States
78712
Indranil.Bardhan@mccombs.utexas.edu

## Abstract

What is the impact of private equity (PE) ownership on healthcare value? Does PE ownership hurt healthcare value, and if so, can its effect be mitigated using health information technologies (IT)? Given the significant investments by PE firms in the healthcare sector in recent years, it is increasingly important for stakeholders, including policy makers, care providers, and patients, to understand their likely impact and whether PE ownership is aligned with their interests. Because PE-owned hospitals often face tradeoffs between achieving short-term, financial returns and pursuing long-term, quality improvement goals, we propose competing hypotheses on how PE ownership may influence healthcare value. More importantly, we argue that IT-enabled health information sharing can mitigate some of these tradeoffs faced by PE-owned hospitals and help them achieve the dual objectives of value-based care: cost reduction and care quality improvement. Using a staggered difference-in-differences approach and data from US hospitals from 2008-2020, we estimate changes in healthcare value following PE ownership. In general, we observe that the overall value of healthcare delivered by hospitals declines after PE ownership. However, our empirical evidence reveals that IT-enabled, health information sharing plays an important moderating role. Hospitals with stronger information-sharing capabilities exhibit greater cost efficiencies and improvements in care quality, leading to higher healthcare value after PE ownership. Furthermore, we find that the type of health information sharing matters. Specifically, we observe that improvements in care quality are primarily driven by information sharing between hospitals and ambulatory care providers, instead of simply hospital-to-hospital sharing of patient health data. We also find evidence that health information sharing facilitates PE-owned hospitals' cost reduction goals by improving workflow efficiency and care quality by reducing adverse events and readmission rates. Our results highlight the critical role of policies and common data standards needed to promote IT-enabled information sharing between healthcare providers, which, in turn, can align the incentives of PE firms with the goals of value-based care.

## 1. Introduction

In recent years, the healthcare sector has witnessed a significant surge in private equity (PE) investments. The total deal value of PE investments in the global healthcare sector exceeded $151 billion in 2021, up from $22.5 billion in 2011 (Biesen and Murphy 2012; Jain et al. 2023). This trend has sparked considerable debate about the impact of PE ownership on the quality of care delivered. Proponents argue that PE firms can provide much-needed capital, improve operational efficiency, and unlock profitability (Borsa et al. 2023). However, critics have raised concerns about the potential negative impact of PE, manifested in the form of poor patient experience (Bruch et al. 2021). Healthcare policymakers, as well as the Federal Trade Commission (FTC) and Antitrust Division of the Department of Justice (DOJ), are concerned that PE may prioritize short-term, cost savings over quality of patient care and patient well-being (Schulte 2022; Hans et al. 2024).

The literature presents mixed evidence on the impact of PE ownership in healthcare. Some studies suggest that PE can lead to improvements in healthcare quality in hospitals and nursing homes (Bruch et al. 2020; Gandhi et al. 2020). However, other studies highlight potential risks, suggesting that hospital care quality declines following PE acquisition (Broms et al. 2024; Gupta et al. 2021; Offodile et al. 2021). This divergence in findings creates significant challenges for policymakers, as it complicates the task of understanding and evaluating the overall impact of PE on healthcare.

The impact of PE on healthcare can be multifaceted and improvements in one area may lead to trade-offs in another. These interdependencies among different dimensions of healthcare performance suggest that a comprehensive evaluation is necessary to fully understand the overall effects of PE ownership, which represents a governance shift that fundamentally alters hospitals' operational priorities, resource allocation, and coordination capacity (Kaplan and Stromberg 2009). In this research, our first research question is stated as follows: *What is the impact of PE ownership on healthcare value?* Following prior literature, healthcare value is defined as the degree to which healthcare providers expend clinical resources effectively to deliver services that improve patient outcomes (Porter 2010; Bardhan et al. 2023). We believe that a focus on healthcare value, often referred to as care value, is especially important in the PE context, as it is aligned with the recent movement toward value-based care (VBC), which emphasizes high-quality and cost-efficient care (Tinetti et al. 2016). Unlike a fee-for-service model, where providers are reimbursed based on the volume of services, the VBC model seeks to optimize patient outcomes while minimizing resource expenditures (Tsevat and Moriates 2018).

Prior studies have highlighted two important benefits of IT-enabled, health information sharing, namely, its effects on reducing operating costs via improved work efficiency and on improving care quality via enhanced care coordination (Ayabakan et al. 2017; Janakiraman et al. 2023; Lammers et al. 2014). Due to the tradeoffs faced by PE-owned hospitals with respect to balancing operating cost reduction and quality improvement, an important question is whether health information sharing can mitigate such tradeoffs due to its dual effects on both cost and quality. Motivated by these observations, our second research question can be stated as follows. *Does IT-enabled health information sharing moderate the impact of PE ownership on care value?*

In our hypothesis development, we argue that PE-owned hospitals could face multiple, often conflicting, objectives: improving care quality, which requires sustained investment and longer time horizons, versus achieving rapid short-term financial returns. As a result, it is difficult to balance short-term financial returns with long-term quality improvement goals, leading to competing hypotheses on how PE ownership could shape healthcare value. On the one hand, as more hospitals adopt VBC programs, achieving higher care value enables hospitals to obtain higher reimbursement rates, which is aligned with PE's financial interests. Hence, we may expect PE ownership to improve care value. On the other hand, research has shown reduction in patient experiential quality and shift toward lower-cost, less-experienced clinical personnel following PE acquisition (Borsa et al. 2023; Bruch et al. 2021; Liu 2022). Hence, declines in care quality may offset or even exceed the efficiency gains derived from cost reduction, thereby hurting care value (Ayabakan et al. 2017; Lake et al. 2020).

The key argument for the unique role of IT-enabled health information sharing in the PE context is that it could help mitigate some of these tradeoffs faced by PE-owned hospitals. Specifically, through its effect on improving workflow efficiency, health information sharing allows PE-owned hospitals to pursue cost reduction without degrading care quality, resulting in higher care value. Meanwhile, through its effect on care coordination, health information sharing also allows PE-owned hospitals to pursue quality improvement without excessively using their own resources and making additional investments, which in turn leads to higher care value.

To test these hypotheses, we combined multiple data sources to construct a panel data of 428 hospitals in the US from 2008 to 2020. Data on PE deals was collected from the SDC Platinum database. We manually matched these PE deals with the affected hospitals through public records. We deployed a staggered difference-in-differences (DID) approach with fixed effects as our empirical approach. Our analysis reveals that compared to non-PE hospitals, hospitals exhibit lower care value following PE ownership. Further analyses of the underlying mechanism suggest that, while the reduction in operating cost increases care value, the overall decline in value is primarily driven by a significant deterioration in care quality. This reduction in care quality is especially driven by lower experiential quality and higher mortality rates after PE acquisition.

More importantly, we find that greater health information sharing can offset the negative impact of PE ownership on care value. Specifically, we find that health information sharing between care providers not only enables PE-owned hospitals to pursue aggressive cost-cutting strategies; it also allows them to provide better care, as measured by lower readmission rates. Interestingly, such an effect is primarily driven by health information sharing between hospitals and ambulatory providers, instead of hospital-to-hospital health information sharing. This is reasonable as this type of information sharing (i.e., exchanging patient data between hospitals and ambulatory providers such as primary care physicians, ambulatory surgery centers, outpatient clinics) can lower readmission rates by facilitating follow-up care, enabling timely intervention, and preventing medication errors (Hesselink et al. 2014).

Our study contributes to the ongoing debate on the role of PE ownership in healthcare organizations. Our study adopts a holistic approach by examining the impact of PE ownership through the lens of care value. This approach is particularly important in the context of the healthcare industry's shift toward VBC models, which reward providers for improving patient outcomes while managing costs (NEJM Catalyst 2017). Our findings indicate that, although PE ownership reduces hospital costs, such cost savings are obtained at the expense of lower care value, raising concerns about alignment between financial and value-based care objectives.

More importantly, we contribute to the literature on the economic and clinical impact of health information sharing between healthcare providers (Ayabakan et al. 2017; Eftekhari et al. 2017; Adjerid et al. 2018; Janakiraman et al. 2023). To the best of our knowledge, this is one of the first studies to demonstrate that health information sharing can help hospitals mitigate the negative impact of PE ownership on care value through both operating cost reduction and care quality improvement. This has important implications for advancing the goals of VBC by showing how technological capabilities can bridge the gap between financial sustainability and care quality.

Our work has important implications for health policy and health IT adoption and use. First, policymakers may find it challenging to design effective regulatory responses without a complete understanding of the overall consequences of PE ownership. By emphasizing a VBC framework, our study highlights the need for a comprehensive assessment of the benefits and risks associated with PE ownership of hospitals. Second, policymakers should consider developing relevant strategic frameworks to ensure greater balance between financial objectives and care value. This could include mandating transparency in post-acquisition quality reporting and enforcing regulations that mandate minimum staffing levels and resources in critical areas. Third, our research highlights the need for healthcare administrators and policymakers to prioritize IT investments that facilitate effective health information sharing, to safeguard against a potential decline in care value after PE ownership.

## 2. Related Literature

### 2.1 Impact of PE Ownership in Healthcare

PE firms bring substantial financial resources and management expertise to healthcare organizations, enabling expansion in service offerings, investments in advanced technologies, and upgrades to facilities that offer the potential to enhance patient care (Robbins et al., 2008). However, PE deals are typically characterized by relatively short investment horizons, usually between three and seven years. This short-term focus may incentivize strategies aimed at maximizing short-term financial gains rather than long-term stability (Richards and Whaley 2024). The growing trend of PE ownership in the healthcare sector has sparked active debate and scholarly interest regarding its implications for healthcare delivery. Researchers have examined the impact of PE ownership in various settings, such as nursing homes (Bos and Harrington 2017; Braun et al. 2020; Braun et al. 2021), hospitals (Bruch et al. 2020; Offodile II et al. 2021; Cerullo et al. 2022; Richards and Whaley 2024), and physician practices (Singh et al. 2022; Bruch et al. 2023). Despite this body of literature, the overall impact of PE ownership on healthcare outcomes remains ambiguous. For instance, Cerullo et al. (2022) reported a reduction in mortality rate after PE acquisition, while Liu (2022) found no significant change. Conversely, Liu (2022) reported a decline in readmission rate, whereas Cerullo et al. (2022) found no effect.

The extant research mostly focuses on the effects of PE ownership on clinical measures of healthcare delivery such as care quality, or financial metrics such as costs. This approach makes it difficult to assess whether PE ownership improved operational performance of healthcare organizations, given the complexity and multi-dimensionality of care delivery. In sum, a notable gap in the literature is the lack of an integrated measure of overall performance resulting from PE ownership. In this research, we seek to address this gap by evaluating the impact of PE ownership from a care value perspective.

### 2.2 Healthcare Value

Healthcare value is a critical concept in the context of measuring healthcare performance. It generally refers to the health outcomes achieved per dollar spent (Porter 2010). This concept emphasizes not just the cost of care but the quality and effectiveness of care provided, and it seeks to optimize both dimensions for a more sustainable healthcare system. Porter and Teisberg (2006) introduced the concept of "value-based competition," emphasizing the importance of measuring outcomes that matter to patients. The Institute for Healthcare Improvement (IHI) also advocates for the "Triple Aim" framework, which focuses on improving patient experience, population health, and on reducing per capita costs (Berwick et al. 2008). It highlights the need for healthcare systems to balance interrelated goals, ensuring that improvements in one area do not come at the expense of another (Bardhan et al. 2023).

Accordingly, we define hospital *care value* as the degree to which a hospital efficiently utilizes its input resources to deliver healthcare services that enhance patient-centered care, i.e. cost-adjusted care quality. This supports a comprehensive assessment of both financial and clinical performance impact of PE ownership on care value delivered by hospitals (Porter 2010). We discuss the operationalization of care value in section 4.2.

### 2.3 IT-enabled Health Information Sharing

The business value of IT is well-documented, with numerous studies highlighting its potential benefits, including increased flexibility, innovation, quality improvement, cost reduction, productivity enhancement, and profitability (Kleis et al. 2012; Melville et al. 2004; Mithas et al. 2017; Pye et al. 2024). In healthcare, health information sharing enables healthcare providers to share patient health data across hospitals, physicians, and laboratories (Adjerid et al. 2018). The literature has highlighted two major benefits of health information sharing. First, by providing timely access to comprehensive patient records, including prior diagnoses, medications, and test results, it reduces unnecessary duplication of tests and procedures and operational costs in labor-intensive processes (Ayabakan et al. 2017). Second, health information sharing also enhances care coordination across providers, improves providers' ability to make well-informed decisions, and contributes to improved care quality (Ayabakan et al. 2017; Janakiraman et al. 2023; Lammers et al. 2014).

Despite these established benefits of health information sharing, how health information sharing shapes the impact of PE ownership on care value is underexplored. Since PE-owned hospitals could face conflicting objectives among different stakeholders, it is critical to study whether IT-enabled health information sharing can mitigate these tradeoffs, which in turn may improve hospital performance, as measured by healthcare value. Our research represents an important step toward answering this question.[1]

## 3. Hypotheses Development

PE ownership represents a distinct form of ownership change in the hospital sector, one that differs substantively from both non-profit and publicly traded, for-profit governance. A significant difference between PE-owned hospitals and non-profit hospitals is that the former are profit-driven and prioritize financial performance, whereas the latter are often community-based and focus on meeting specific social or mission-driven objectives (Horwitz and Nichols 2022). Unlike publicly traded, for-profit hospitals, which balance shareholder returns with ongoing obligation to long-term reputational concerns, PE firms operate under much shorter investment horizons (typically 3–7 years) with an explicit mandate to generate high returns upon exit through resale or IPO (Offodile II et al. 2021; Richards and Whaley 2024). This timeline creates stronger incentives for rapid cost restructuring, labor optimization, and financial engineering compared to other for-profit hospitals (Konda et al. 2019).

[1] Besides health information sharing among care providers, a hospital's IT capabilities also include its internal IT capabilities that support within-hospital functions such as electronic clinical documentation and computerized provider order entry. However, as shown in the literature, the benefits of internal IT are often constrained if information cannot be shared across organizational boundaries (Walker et al. 2005; Atasoy et al. 2019). That is, it does not address care coordination capabilities across hospitals. Hence, our focus is on inter-hospital information sharing although we control for hospitals' internal IT capabilities in our analyses.

Together, these characteristics position PE ownership as a unique governance structure that could create tradeoffs among multiple goals and different stakeholders. In this section, we will hypothesize the implications of PE ownership on care value delivered by hospitals and how IT-enabled health information sharing may alleviate some of these tradeoffs and thereby shape healthcare value.

### 3.1 Impact of PE Ownership on Care Value

We draw on the resource-based view (RBV) theory of the firm to understand the impact of PE ownership on healthcare value. RBV theory emphasizes that firms create value or maintain sustainable competitive advantage by acquiring, developing, and effectively deploying valuable, rare, inimitable, and non-substitutable resources and capabilities (Peteraf 1993). These resources and capabilities can be tangible, such as physical assets and human capital, or intangible, including managerial experience, organizational routines, and reputation. In the context of PE ownership, hospitals are often restructured to better leverage their internal resources and capabilities in ways that drive performance improvement and financial returns. Further, PE firms typically bring in external managerial expertise, performance monitoring systems, and financial oversight, that allow strategic deployment of resources (Meuleman et al. 2009). These intangible assets can improve decision-making, accountability, and resource allocation. From an RBV perspective, this reconfiguration of operational activities and organizational resources may allow for better alignment with care value creation.

More importantly, as an increasing number of hospitals adopt VBC programs, enhancing care value makes it easier for hospitals to secure higher reimbursement rates. These VBC incentives directly tie financial incentives to improvements in care value. Because PE firms are strongly motivated to enhance the financial performance of portfolio hospitals, gains in care value can translate into higher margins and greater exit valuations (Powers et al. 2021). In this sense, improvements in care value are closely aligned with PE's financial interests. Hence, PE firms may have a strong incentive to implement initiatives that can improve care value within their acquired hospitals.

*Hypothesis 1(a): PE-owned hospitals exhibit higher care value than non-PE-owned hospitals.*

However, PE-owned hospitals also face multiple, often conflicting objectives: improving care quality, which requires sustained investment and a longer time horizon, versus achieving short-term financial returns. PE firms are primarily accountable to investors and operate under specific incentives designed to maximize returns within a limited time frame (Kaplan and Stromberg 2009). These tensions make it difficult to balance short-term, financial returns with long-term, quality improvement goals.

Consistent with this tension, the literature documents staffing reductions and shifts toward less-experienced clinical personnel following PE acquisition of hospitals (Borsa et al. 2023). Meanwhile, the literature has also reported a decline in patient experiential quality and other clinical outcomes following PE acquisition (Braun et al. 2021; Bruch et al. 2021; Gupta et al. 2021; Liu 2022). Furthermore, because PE acquisitions are often structured as leveraged buyouts, substantial debt service obligations can constrain operating budgets, further limiting resources available for staffing, clinical operations, and quality improvement (Gupta et al. 2021).

Although these restructuring efforts may produce short-term cost savings, a decline in care quality may offset or even exceed the efficiency gains derived from cost reduction. In this case, PE ownership can reduce hospitals' ability to effectively utilize resources to generate better patient health outcomes, thereby diminishing healthcare value. Hence, we propose a competing hypothesis.

*Hypothesis 1(b): PE-owned hospitals exhibit lower care value than non-PE-owned hospitals.*

### 3.2 Moderating Role of Health Information Sharing

#### 3.2.1 Operating Cost Reduction

Stakeholder theory highlights that businesses operate within a diverse ecosystem, where different stakeholders have distinct goals and expectations, and that long-term success depends on the well-being of all stakeholders (Donaldson and Preston 1995; Freeman and McVea 2001; Parmar et al. 2010). In the context of hospitals, this includes delivering better patient health outcomes, ensuring favorable working conditions for staff, and advancing the broader mission of high-quality, patient-centered care, which are critical stakeholder concerns that extend beyond financial returns.

A common strategy of PE-owned hospitals is operating cost reduction, particularly labor costs, which comprise roughly 60% of total hospital expenses (AHA 2024). Reducing labor costs offers an effective, short-term approach to enhance profitability and improve financial performance (Borsa et al. 2023). Moreover, many PE transactions are structured as leveraged buyouts, where a substantial portion of the purchase is financed through debt, using the acquired hospital's assets as collateral (Cai and Song 2023). This financial structure increases pressure to generate higher returns to meet debt obligations and investor expectations. As a result, PE firms are motivated to cut operating costs in the acquired hospitals to maintain cash flow and preserve margins.

However, a focus on operating cost reduction may conflict with the core mission of hospitals to prioritize community health and patient outcomes over financial considerations (Meliones 2000). Hospitals tend to maintain slack resources and adequate staffing levels to ensure service quality. Many care providers lack clear understanding of operating costs, making it difficult to work with PE firms to identify opportunities to improve cost efficiency, reduce delays, or eliminate non-value-adding activities (Porter and Lawrence 2011). From a stakeholder theory perspective, PE's cost-reduction goal may be misaligned with hospitals' goals for clinical continuity, adequate staffing, and patient-centered service.

We argue that IT-enabled health information sharing between care providers can reconcile conflicting goals among PEs and hospitals. Specifically, health information sharing allows PE-owned hospitals to implement efficient workflows, and thus, pursue cost reduction without degrading care quality. First, IT-enabled health information sharing facilitates automation, enabling hospitals to reduce reliance on labor-intensive processes (Adjerid et al. 2018; Bardhan et al. 2023). Second, timely access to critical patient information including laboratory tests, medical histories, and imaging reports enables clinicians to reduce the need for duplicate tests and procedures (Ayabakan et al. 2017; Eftekhari et al. 2017). Third, once the IT infrastructure for health information sharing is in place, it can be scaled efficiently, as sharing additional patient records typically entails automated electronic processes that do not require significant additional manual labor or cost (Torab-Miandoab et al. 2023).

In sum, from a stakeholder theory perspective, health information sharing helps mitigate the tension between the cost reduction goal of PE-owned hospitals and their patient-centered care mission by improving workflow efficiency. This alignment enables PE-owned hospitals with high levels of information sharing to reduce operating costs without compromising quality, compared to those with low health information sharing. When care quality is not compromised, such operating cost reductions could, in turn, improve care value, i.e., cost-adjusted quality. Accordingly, we propose the following hypothesis.

*Hypothesis 2: Health information sharing positively moderates the impact of PE on care value through its role in facilitating operating cost reduction.*

### 3.2.2 Care Quality Improvement

PE-owned hospitals also have incentives to improve care value through better clinical quality. First, as noted earlier, they face increasing pressure to improve clinical outcomes due to the widespread shift toward VBC, which ties reimbursements to performance on quality (Chernew et al. 2011; VanLare and Conway 2012). Programs such as the Hospital Readmissions Reduction Program (HRRP), for example, financially penalize hospitals with high readmission rates (Kripalani et al. 2014). Thus, making quality improvement, especially readmission metrics, can be financially beneficial and contribute to margins for PE-owned hospitals.

Second, PE-owned hospitals operate in a competitive marketplace where patient choice and hospital reputation significantly influence demand (Noether 1988; Kessler and McClellan 2000; Mukamel et al. 2002). Patients are sensitive to hospital quality indicators such as readmission rates (Luft et al. 1990). Varkevisser et al. (2012) reported that a one percentage reduction in readmission rates was associated with a 12% increase in hospital demand. Hence, enhancing care quality can not only increase revenue in PE-owned hospitals but also improve market share. This is particularly relevant given the short-term, investment horizon of PE firms, which aims to increase the value of portfolio hospitals before pursuing an exit (Richards and Whaley 2024).

However, quality improvement could conflict with the short-term, investment horizon of PE firms. Enhancing quality typically requires additional investments in staffing, training, equipment, and coordination processes (Endalamaw et al. 2024; Needleman et al. 2002), whose returns may materialize over a longer horizon (Thusini et al. 2022). Nevertheless, the short-term investment horizon requires PE to extract returns within a short timeframe (Kaplan and Stromberg 2009). Due to these conflicting objectives, PE-owned hospitals may not be able to effectively improve care quality.

We argue that IT-enabled health information sharing can mitigate this tension by allowing PE-owned hospitals to pursue quality improvement without excessively utilizing their own resources and making additional investments. Specifically, health information sharing provides hospitals with stronger coordination and information-processing capabilities, by enabling timely access to external patient records, test results, and treatment history (Bardhan et al. 2023). Prior research shows that such coordination improves patient care quality by supporting clinical decision-making, reducing errors and adverse events, and improving patient safety (Smith et al. 2005; Walker et al. 2005; Haque et al. 2020; Bao and Bardhan 2022; Chiedi 2019; Janakiraman et al. 2023). It also facilitates better provider-provider communication about patient care transitions, thereby streamlining the referral process in complex settings (Eftekhari et al. 2023; Fecher et al. 2021). These benefits of health information sharing can translate into effective improvements in patient readmission rates that are largely influenced by clinical decisions and care coordination.

Therefore, we argue that PE-owned hospitals with greater information sharing capabilities can better align PE's care quality improvement goals with their short-term financial interests, compared to hospitals with low health information sharing. By improving coordination and clinical responsiveness, hospitals with high information sharing capabilities are more capable of improving care quality without substantial additional spending in resources, thereby contributing to greater care value. Accordingly, we propose the following hypothesis.

*Hypothesis 3: Health information sharing positively moderates the impact of PE on care value through its role in facilitating care quality improvement.*

## 4. Data and Variables

### 4.1 PE Deals and Affected Hospitals

We focus on hospitals involved in leveraged buyouts (LBOs), a common form of PE deals. Our primary data on LBO deals comes from the PE and M&A databases of SDC Platinum. Instances where PE firms provide debt financing or purchase minority stakes are excluded, because they do not provide the same level of operational and strategic control as LBO deals. To identify PE transactions in the healthcare sector from the PE database, we followed the following approach: (1) Select the "Investee Company Nation" as United States and "Investment Date" between 2000 and 2024, (2) Restrict "Investee SIC" to general medical and surgical hospitals, and (3) Require the "Investee Company Long Business Description" to contain the word "hospital". We supplement the main LBO deals with M&A deals in our sample where the acquirer is a PE company or PE fund. To compile a list of PE transactions in the healthcare sector from the M&A database, we adopted the following procedures: (1) Select the "Target Nation" as US and "Date Announced" between 2000 and 2024, (2) Restrict "Deal Status" to completed and "Acquisition Techniques" to leveraged buyout, and (3) Restrict "Target SIC" as general medical and surgical hospitals. To focus specifically on PE deals, we require "Acquiror Immediate Parent Public Status" and "Acquiror Ultimate Parent Public Status" to be private.

After identifying PE deals in US hospitals, for deals at the level of a hospital system, we used archived webpages from the Wayback Machine to confirm the hospital system's each hospital name, location, ownership structure, and affiliation at the time of acquisition. When the webpage information was unavailable in the Wayback Machine, we supplemented it by conducting targeted searches of press releases announcing the acquisition or reviewing the hospital system's annual financial filings (e.g., 10-K) for the corresponding year of the deal. These reports provided organizational structure and hospital lists, allowing us to confirm whether individual hospitals were included in each system-level transaction. Overall, the manual verification process resulted in complete ownership and location information for all hospitals in the PE treatment group, which led to the final sample of 115 PE-owned hospitals for our analysis. We then merged this data with other hospital-level data from the Centers for Medicare and Medicaid Services (CMS) Cost Reports, staffing levels from the American Hospital Association (AHA) survey, and quality data from the CMS Hospital Compare database to measure other variables, as discussed in greater detail.

It is important to note that if a hospital system experienced multiple PE deals, we only consider the first deal. This approach reflects the fact that PE ownership is essentially an absorbing state, i.e. PE ownership would fundamentally change hospitals' operational priorities, resource allocation, and coordination capacity (Kaplan and Stromberg 2009). We also verified that none of these PE-owned hospitals in our sample reverted to non-PE ownership through PE exits.

**4.2 Healthcare Value**

Quantifying healthcare value is challenging, as it involves measuring multiple inputs and outputs that are often interdependent and complex. To address this challenge, we adopt a Data Envelopment Analysis (DEA) approach. It is a nonparametric optimization method that evaluates how efficiently a decision-making unit (DMU) converts its input resources into clinical outputs, i.e., a hospital in our case. DEA has been widely applied in healthcare to measure hospital efficiency because it accommodates multiple inputs and outputs without requiring strong parametric assumptions (Ayabakan et al. 2017; Bardhan et al. 2023; Tiemann and Schreyögg 2012).

DEA constructs a Pareto-efficient frontier across all hospitals and computes a radial efficiency score $\theta$ for each hospital. Hospitals on this frontier receive a DEA score of one, indicating optimal efficiency. Hospitals below the frontier receive scores between zero and one, reflecting each hospital's proportional distance from the frontier and its relative efficiency compared to the most efficient hospitals. *Care Value* is thus measured based on the DEA score $\theta$, with higher values indicating higher efficiency, i.e. greater healthcare value.

Following Bardhan et al. (2023), we apply an input-oriented Banker-Charnes-Cooper (BCC) model (Banker et al. 1984) as specified in equation (1).

$$\begin{aligned} &\text{Min}_{\theta,\lambda}\{\theta - \varepsilon(1's^- + 1's^+) \mid X\lambda + s^- = \theta x_o; \\ &Y\lambda - s^+ = y_o; 1'\lambda = 1; \lambda \geq 0\} \end{aligned} \tag{1}$$

*X* and *Y* denote the input and output vectors of the focal hospital, respectively, and θ represents the DEA-based score of healthcare value. Reflected in the constraint, $1'\lambda = 1,$ this model allows for variable returns to scale, as hospitals typically exert greater control over their resources input utilization than over outputs.[2]

---

[2] We also tested the robustness of these results using bootstrapped DEA (Simar and Wilson 1998), which yields consistent estimates. These results are available upon request.

To evaluate hospital-level, healthcare value, our DEA model utilizes operating expenses, capital expenses, and labor (including physicians, nurses, and other clinicians) as inputs, and patient experiential quality, mortality rate, and readmission rate as outputs. Experiential quality is measured based on an average score of five survey questions on patient satisfaction with communication and responsiveness of hospital staff (Senot et al. 2016).[3] Mortality rate is computed as the weighted average of individual mortality rates for heart attack, heart failure, and pneumonia, with the weight based on the number of patients with each condition. Similarly, hospital readmission rate is calculated as the weighted average of readmission rates for these three conditions, respectively. To account for differences in patient volume across hospitals, we normalize input resources based on the number of encounters, resulting in inputs measured on a per-encounter basis. Due to the structure of the DEA model, we use the inverse of mortality and readmission rates as output variables, because lower rates of mortality and readmission indicate higher care quality.

### 4.3 Health Information Sharing and Health IT Capabilities

We construct the health information sharing variable *InfoShare* using a set of questions from the AHA IT Supplement survey. These questions evaluate the extent to which hospitals electronically exchange or share different types of patient health data with various healthcare providers. The survey measures information sharing across five key categories of patient health data: (a) patient demographics, (b) clinical care records, (c) laboratory results, (d) medication history, and (e) radiology reports. For each category, the survey asks whether the hospital shares data with four types of providers: (a) hospitals within the same health system, (b) hospitals outside the health system, (c) ambulatory providers within the system, and (d) ambulatory providers outside the system. This yields a total of 20 survey responses. For each question, hospitals indicate whether they electronically share the specified data type with the specified provider type. Hence, *InfoShare* is calculated as the percentage of "Yes" responses across all questions at the hospital-year level. A higher value of *InfoShare* reflects more health information sharing between the focal hospital and other care providers. The specific questions used to construct *InfoShare*

[3] Patient satisfaction data are drawn from the HCAHPS survey in the Hospital Compare database, including measures such as nurse communication, doctor communication, staff responsiveness, communication about medications, and communication about post-discharge instructions.

are described in the Online Appendix Table OA1.

In addition to *InfoShare*, we also test the heterogeneous effects of hospital-to-hospital and hospital-to-ambulatory provider information sharing. We distinguish between these two dimensions of health information sharing by measuring each type separately. Hospital-to-hospital information sharing (*InfoShare_Hos*) refers to the exchange of patient health data with other hospitals. Hospital-to-ambulatory provider information sharing (*InfoShare_Amb*) measures the level of data exchange between the focal hospital and ambulatory providers, which include physician offices, ambulatory surgery centers, and outpatient clinics.

To account for differences in the overall health IT infrastructure between hospitals affected by PE ownership and those that are not, we follow the literature on IT capabilities (Pye et al. 2024; Rai and Tang 2010) and construct two variables, *Intra-hospital IT* and *Inter-hospital IT,* for matching purposes. *Intra-hospital IT* refers to a hospital's internal ability to collect, process, and use clinical and administrative information across departments within the same facility for care delivery. It is measured based on a set of questions that assess the focal hospital's IT capabilities in several areas: (a) electronic clinical documentation, (b) results viewing, (c) computerized provider order entry (CPOE), (d) clinical decision support, and (e) other functionalities of the EHR system. Each of these areas is comprised of multiple questions, as shown in section 1 of Online Appendix Table OA1. *Intra-hospital IT* is calculated as the percentage of responses indicating whether the corresponding capability is either fully implemented in at least one unit or across all units within the focal hospital.

*Inter-hospital IT* is defined as the broader IT capability to digitally share patient information with external entities, including both care providers and patients. We use two sets of questions in the AHA IT supplement survey to measure this construct. The first set of questions provides a measure of *InfoShare*. The second set assesses whether patients at the focal hospital can perform various tasks online, as shown in section 3 of Online Appendix (Table OA1). *Inter-hospital IT* is calculated as the percentage of "Yes" responses across the two sets of questions at the hospital-year level.

**4.4 Control Variables**

To mitigate concerns related to omitted variable bias, we control for a set of hospital-specific characteristics that may influence both the likelihood of being a target for PE ownership and hospital performance. Specifically, we include the log of the average daily census, log of total inpatient days, case mix index, outlier adjustment factor, log of the total number of beds, level of Intra-hospital IT, resident-to-bed ratio, and wage index. These control variables collectively account for patient volume, service complexity, hospital size, health IT infrastructure, and other hospital-specific characteristics that may be important predictors of hospital performance.

## 5. Empirical Analyses

### 5.1 Baseline Model Specification

To test whether PE ownership improves or hurts care value (H1a & H1b), we utilize a staggered difference-in-differences (DID) approach. Hospitals affected by PE ownership are classified in the treatment group. Since the timing of PE ownership varies across hospitals, the post-treatment period differs for each hospital in the treatment group. Other hospitals that are not affected by PE ownership are classified in the control group. The baseline model is specified in equation (2), with the unit of analysis at the hospital (*i*) – year (*t*) level.[4]

$$Care\ Value_{it} = \lambda_0 + \lambda_1 Treatment_i + \lambda_2 Post_{it} + \lambda_3 Post_{it} * Treatment_i + \gamma X_{it} + u_i + \eta_t + \varepsilon_{it} \quad (2)$$

The dependent variable $Care\ Value_{it}$ represents the care value of hospital $i$ in year *t*. It is a continuous variable ranging from zero to one, where higher values indicate greater care value. The independent variable $Post_{it} * Treatment_i$ indicates the PE ownership status of hospital $i$ in year *t*, taking a value of one if hospital $i$ experienced PE acquisition in year *t* and zero otherwise. $Post_{it}$ indicates whether the year *t* belongs to the year of post-treatment period for both treated and control hospitals. The vector $X_{it}$ includes a set of time-varying control variables. It includes variables that represent hospital $i$'s patient volume and complexity in year *t,* such as log of the average daily census, log of total inpatient days, case mix index, and outlier adjustment factor. Further, we control for characteristics of the hospital's operating environment, such as its number of staffed beds, resident-to-bed ratio, and wage index for hospital *i* in year *t*. We also include a control for overall IT capabilities, measured by the level of *Intra-hospital IT* in hospital *i* in year *t*. We include both hospital and year fixed effects, denoted as $u_i$ and $\eta_t$ to account for unobserved, time-invariant differences at the hospital level and general time trends, respectively. $\varepsilon_{it}$ represents the error term.

[4] Although we include the term $Treatment_i$ in the model equation (2), its effect is absorbed by the hospital fixed effect $u_i$. As a result, it is automatically dropped from the regression, and we are unable to estimate $\lambda_1$.

### 5.2 Moderating Effect of Health Information Sharing

To explore the moderating role of health information sharing on the relationship between PE ownership and care value (H2 & H3), we deploy the following model specification (3).[5]

$$\begin{aligned} Care\ Value_{it} = {} & \lambda_0 + \beta_1 Post_{it} + \beta_2 Treatment_i + \beta_3 Post_{it} * Treatment_i + \beta_4 Post_{it} * InfoShare_i \\ & + \beta_5 Treatment_i * InfoShare_i + \beta_6 Post_{it} * Treatment_i * InfoShare_i \\ & + \beta_7 InfoShare_i + \gamma X_{it} + u_i + \eta_t + \varepsilon_{it} \end{aligned} \tag{3}$$

In this model, we include interaction terms $Post_{it} * InfoShare_i$, $Treatment_i * InfoShare_i$, and $Post_{it} * Treatment_i * InfoShare_i$ while holding other variables the same as in model (2). $InfoShare_i$ is a continuous measure of the level of health information sharing by the focal hospital before PE acquisition (i.e., one year pre-treatment). This measure allows us to address potential endogeneity concerns due to reverse causality. As discussed in Section 4.3, $InfoShare_i$ takes a value from zero to one, with higher values indicating higher level of health information sharing one year preceding the treatment year. We also construct an instrumental variable for $InfoShare_i$ to address the endogeneity concerns related to this variable, as discussed in greater detail in section 5.4.

### 5.3 Matching

While our analysis includes a group of covariates as controls, unobserved time-varying confounding factors may lead to potential endogeneity concerns. To strengthen the validity of our DID analysis and address potential selection bias, we implement a one-to-three propensity score matching (PSM) approach to construct a comparable control group. This method allows us to account for pre-treatment differences and ensure parallel trends in care value between PE-treatment hospitals and hospitals in the control group (Heckman et al. 1998). Specifically, we match each treated hospital with three nearest neighbor control hospitals without replacement, based on the average values of key hospital characteristics in the pre-treatment period. These matching variables include the number of physicians and nurses, which represents the scale and flexibility of the hospital's clinical workforce; experiential quality, readmission rate, mortality rate, and care value, which jointly capture care delivery performance in terms of quality and efficiency; and the level of intra- and inter-hospital IT infrastructure, which reflect hospital's digital capability that may influence PE ownership. This yields 115 hospitals in the treatment group and 319 hospitals in the control group, resulting in 5,096 hospital-year observations during the study period from 2008 to 2020. [6]

[5] In equation (3), the terms $Treatment_i$, $Treatment_i * InfoShare_i$ and $InfoShare_i$ are dropped due to the hospital fixed effect $u_i$. Thus, $\beta_2$, $\beta_5$ or $\beta_7$ are not reported in Tables 4, Tables 6 and Table 7.

Table 1 presents descriptive statistics for the matched sample, along with balance test results comparing treatment and control hospitals prior to PE acquisition. The table is organized into three panels. Panel A (*DEA Variables*) lists all input and output variables, as well as the estimated care value scores. Panel B (*Control Variables*) includes the covariates in vector $X_{it}$. Panel C (*Other Variables*) reports additional variables used either in propensity score matching or as dependent variables in supplementary analyses. According to the p-values in the last column, most of the listed variables do not differ significantly between treatment and control groups. The results indicate that the matching procedure achieves balance not only for variables used in the PSM algorithm, but also among covariates not explicitly included in the matching process. This enhances the credibility of our identification strategy and increases confidence that observed post-treatment effects can be more reliably attributed to the effects of PE ownership rather than to pre-existing differences between the two groups. We use this matched sample throughout all analyses presented in the paper.

### 5.4. Endogeneity and Selection Bias

As noted above, an important concern about our identification strategy in estimation of equation (2) is that PE acquisition decisions are endogenous, as they may be driven by the strategic selection of PE firms. We utilize PSM to address endogeneity concerns arising from observable covariates that can influence both the likelihood of PE acquisition and hospital outcomes. However, concerns may still exist due to potential endogeneity arising from complex interactions or nonlinear relationships among covariates, which may not be fully captured by parametric models (e.g., logistic regressions) typically used in PSM. To address such concerns, we employ the double machine learning (DML) method. DML is a non-parametric machine learning approach that can address endogeneity issues arising from high-dimensional confounders or their effects on the treatment selection and hospital outcomes, which cannot be satisfactorily modeled using parametric methods (Chernozhukov et al. 2018). Furthermore, to address potential endogeneity concerns related to the level of health information sharing in equation (3), we apply an instrumental variable (IV) approach. In the following section, we discuss our DML and IV approaches, respectively.

[6] We also implemented 1:1 Coarsened Exact Matching (CEM) using the same set of matching variables (Iacus et al. 2012), resulting in 109 treated hospitals matched to 109 controls. The estimated effects of PE ownership on care value and its underlying effects on cost and care quality are consistent with our baseline results.

**5.4.1 Double Machine Learning**

To address potential identification concerns in our DID model (i.e., selection bias in treatment assignment), we employ a DML approach which provides a more flexible framework to control for covariates that may affect both treatment selection and outcomes (Chernozhukov et al. 2018). Following Xu et al. (2024), we adopt a three-step DML procedure to estimate the effect of PE ownership on care value (i.e., $\lambda_3$ in equation (2)).

$$\text{Step1: } Care\ Value_{it} = g_0(\boldsymbol{Z}_{it}) + \varphi_i + \tau_t + \epsilon_{it} \tag{4}$$
$$\text{Step 2: } Treatment_i * Post_{it} = m_0(\boldsymbol{Z}_{it}) + \varphi_i + \tau_t + \mu_{it} \tag{5}$$
$$\text{Step 3: } \left(Care\ Value_{it} - \widehat{Care\ Value_{it}}\right) = \delta \times \mu_{it} + \gamma_{it} \tag{6}$$

where $Care\ Value_{it}$ represents the key dependent variable in our main analyses, and $\varphi_i$ and $\tau_t$ are the hospital and time dummies, respectively. $\epsilon_{it}$ is the error term in step 1 which represents the portion of the outcome unexplained by the control variables and fixed effects. $\mu_{it}$ is the residual in step 2, representing the unexplained portion of the treatment assignment. δ is the key parameter of interest in step 3, which represents the causal effect of PE ownership on care value.

The DML approach follows a three-step procedure. In step 1, we model the outcome variable (i.e., care value) as a function of $\boldsymbol{Z}_{it}$ and hospital and year dummies. Vector $\boldsymbol{Z}_{it}$ represents a set of variables including the controls in our DID model and variables affecting the PE market (PE M&A secondary sale, and IPO), which may influence a PE firm's decision to invest in the local hospital market. Specifically, to measure PE market characteristics, we include the deal volume and deal value of PE exit with mergers and acquisitions, secondary sales, and initial public offerings (IPOs) that occurred within the same Core-Based Statistical Area (CBSA) over the past two years.[7] In step 2, we model the treatment term using the same variables as in step 1. The functions $g_0(\cdot)$ and $m_0(\cdot)$ in equations (4) and (5) are estimated using machine learning. In our DML analysis, we use gradient boosted tree, a non-parametric model, for both $g_0(\cdot)$ and $m_0(\cdot)$.

The key intuition behind this three-step DML process is that PE acquisition decisions may be correlated with observable characteristics of hospitals and their local PE markets. Simply regressing the outcome on the treatment variable without isolating these confounding factors would introduce selection bias in the estimated effect of PE ownership. By separately modeling and removing the explainable components of both the outcome and treatment assignment (steps 1 and 2) and then regressing the residualized outcome on the residualized treatment indicator in step 3, DML yields a more robust estimate of the treatment effect. This procedure effectively "doubles" the use of machine learning to isolate the causal impact of PE ownership on hospital care value.

**5.4.2 Instrumental Variable Estimation**

A critical concern about the identification strategy in the estimation of equation (3) is that omitted variables may be correlated with both hospital care value and the level of health information sharing. For instance, hospitals that are owned by larger health systems may have better capability to share health information and be more likely to deliver higher care value. To address this endogeneity concern, we follow the approach adopted in the literature and use the level of information sharing of neighboring hospitals as the IV for $InfoShare_i$ (Lu et al. 2018; Sun et al. 2020). Specifically, based on the AHA data, we first identify the neighboring hospitals located in the same hospital referral region (HRR) for each focal hospital. We calculate the average level of health information sharing across neighboring hospitals, one-year preceding treatment, denoted as $NeighborInfoShare_i$, using a similar approach as described in section 4.3. Then we use this average value as the IV for $InfoShare_i$. As discussed in greater detail later, we also create IVs for hospital-to-hospital information sharing (*InfoShare_Hos*) and hospital-to-ambulatory provider information sharing (*InfoShare_Amb*), denoted as

[7] The CBSA-level PE exist information is collected from the PE Exist database of SDC Platinum.

$NeighborInfoShare_Hos_i$ and $NeighborInfoShare_Amb_i$ respectively.

Miller and Tucker (2009) suggest a network effect where the adoption of health IT by one hospital is influenced by the adoption decisions of other hospitals within the same region. Dranove et al. (2014) highlight the role of shared local IT services in shaping the health IT adoption patterns within local markets. Building on these insights, we argue that greater health information sharing at neighboring hospitals increases the likelihood of the focal hospital to also participate in health information sharing. In other words, this IV is valid and satisfies the relevance condition.

Furthermore, we argue that the exclusion restriction is satisfied for the following reasons. First, hospital care value is primarily influenced by its internal operations, including staffing, clinical processes, and resource utilization. While neighboring hospitals' adoption of health information sharing may influence broader regional trends, it is unlikely to directly affect patient flow or care delivery processes at the focal hospital. In other words, information sharing at neighboring hospitals is unlikely to have a direct impact on patient care at the focal hospital.

We conduct several falsification tests to examine the association between our IV and care value, patient flow, and staffing levels at the focal hospital. The results of exogeneity tests are reported in Table A1 of the Online Appendix. We observe that the IV is not directly associated with care value, number of inpatient visits, inpatient days, or labor at the focal hospital. This suggests that information sharing of neighboring hospitals is unlikely to be correlated with changes in the focal hospital's operations that may impact its care value. Overall, our IV satisfies both relevance and exogeneity conditions. Hence, we use this IV in a two-stage least-square (2SLS) regression of equation (3) to estimate the heterogeneous effect of PE ownership on care value for hospitals with different levels of *InfoShare*.

## 6. Results

### 6.1 Impact of PE Ownership on Care Value

Table 2 shows the baseline estimate of the impact of PE ownership on hospital care value. The results using care value as the dependent variable are reported in column (1), followed by the results in columns (2) to (7) which use the DEA inputs and outputs as separate dependent variables. The coefficient estimate of *Post*×*Treatment* in column (1) suggests that PE-owned hospitals exhibit care value that is 0.034 (i.e., 10.9%) lower compared to non-PE hospitals.[8] The DML results are presented in Table A2 and are consistent with the main DID finding.

The results in columns (2) to (4) suggest that PE-owned hospitals exhibit lower operating costs, non-labor operating costs, and labor FTE, respectively. For instance, the impact of PE ownership reduces average hospital operating costs by almost 20 million dollars (coeff. = -19.995, p-value < 0.01). However, as observed in columns (5) and (6), the decline in care value is driven primarily by a decline in care quality. Specifically, we observe a 0.788 (i.e., 1.08%) reduction in experiential quality and 0.266 (i.e., 2.02%) increase in mortality rate after PE acquisition. The result in column (7) suggests that readmissions are not significantly affected by PE ownership. A plausible explanation is that readmission rates are closely monitored and penalized under the CMS Hospital Readmissions Reduction Program (HRRP). Hence, although cost-cutting strategies can lead to an increase in readmission rates, PE-owned hospitals face strong incentives to prevent deterioration in this metric to avoid financial penalties.

Overall, the findings highlight a critical trade-off: while operating cost reduction improves efficiency and increases care value, there are significant declines across a range of dimensions of care quality following PE acquisition, which decreases care value. These results highlight that PE ownership, while enhancing hospital financial performance, may inadvertently erode the core elements of care value that are most crucial to patient well-being.

The validity of our DID estimation relies on the assumption of parallel trends. To assess whether treated and control hospitals exhibit parallel trends in care value prior to the PE acquisition, we employed a leads/lags model (Autor, 2003), as specified in equation (7).[9]

$$\begin{aligned} Care\ Value_{it} = {} & \lambda_0 + \lambda_1 Treatment_i + \sum_{\tau=-5}^{-1} \lambda_\tau Pre_{it+\tau} + \lambda_0 Post_{it+0} + \sum_{\tau=1}^{5} \lambda_\tau Post_{it+\tau} \\ & + \sum_{\tau=-5}^{-1} \beta_\tau Pre_{it+\tau} * Treatment_i + \beta_0 Post_{it+0} * Treatment_i \\ & + \sum_{\tau=1}^{5} \beta_\tau Post_{it+\tau} * Treatment_i + \gamma X_{it} + u_i + \eta_t + \varepsilon_{it} \end{aligned} \tag{7}$$

[8] We also studied potential heterogeneity effects of PE investment experience and observe that hospitals that are acquired by high-experience PE firms are more likely to exhibit greater care value and experiential quality, and higher operating costs, compared to hospitals owned by low-experience PE firms.

[9] Like equation (2), the term $Treatment_i$ is dropped due to the hospital fixed effect $u_i$.

Our sample consists of hospitals with data that spans ten pre-treatment and/or post-treatment years. For the sake of space and clarity, we group every two years to construct the $Pre_{it+\tau}$ and $Post_{it+\tau}$ variable, where τ represents the relative time period (in two-year intervals) before or after the PE acquisition for each hospital.[10] Specifically, for periods prior to treatment ($\tau < 0$), the variable $Pre_{it+\tau}$ represents an indicator variable that equals one if year $t$ is $-2\tau$ or $-2\tau - 1$ years before the treatment start year. For example, $Pre_{it-1}$ indicates that, the focal year *t* is either 1 or 2 years before the treatment start year for hospital *i*. Similarly, for periods after treatment ($\tau > 0$), $Post_{it+\tau}$ represents an indicator variable that equals one if year *t* is $2\tau$ or $2\tau - 1$ years after the treatment start year for hospital *i*. For example, $Post_{it+1}$ indicates that the focal year *t* is either 1 or 2 years after the treatment start year for hospital *i*. $Post_{it+0}$ indicates that year *t* is the treatment start year for hospital $i$. For hospitals with more than 8 post-treatment years, we collapse the subsequent years into a period labeled as "4+". For example, we combine "$Post_{it+4}$" and "$Post_{it+5}$" to form a new dummy, denoted as "$Post_{it+(4+)}$". To ensure consistency in comparisons across time periods, we designate $Pre_{it-5}$ and the interaction term $Pre_{it-5} * Treatment_i$ (representing 9 or 10 years before the treatment year) as the comparison period. All other time periods, both pre-treatment and post-treatment, are evaluated against this baseline.

The leads/lags regression results are presented in Table A3. The coefficients corresponding to the pre-treatment periods, ranging from *Post(-4)×Treatment* to *Post(-1)×Treatment* are statistically insignificant, indicating that differences between the treated and control hospitals in the years prior to the PE acquisition are not significant. This result supports the parallel trend assumption. In contrast, the coefficients for post-treatment periods, denoted by the terms *Post(1)×Treatment* to *Post(4+)×Treatment* are mostly statistically significant. Further, the increasing magnitude of the coefficients suggests that the negative impact of PE ownership on care value becomes more pronounced over time.

To further validate our DID estimation approach, we conduct placebo tests of spurious trends by randomly assigning treated hospitals. Specifically, we randomly assign treated hospitals to form the treatment group by selecting 115 hospitals from our main estimation sample (i.e., 115 out of 428). We then estimate the DID regression on the randomly assigned treatment group and repeat this process 1,000 times, resulting in a total of 1,000 experiments and 1,000 regression results. We graph a kernel density plot of the results as shown in Figure A1 of the Appendix. Specifically, the X-axis represents the estimated treatment effects and the Y-axis represents the probability densities and the respective p-values associated with the coefficient estimates in each experiment. The distribution is centered around zero, since the treatment was randomly assigned. This serves as a validity check, ensuring that the DID estimator does not systematically generate spurious treatment effects. We also report a rejection rate of 0.065 for the null hypothesis of no treatment effect at a significance level of 5% ($p < 0.05$). This rate is close to the expected 5% threshold, confirming the validity of our DID analysis.

[10] The results based on yearly measurement of $Pre_{it+\tau}$ and $Post_{it+\tau}$ are consistent and available upon request.

Moreover, the econometrics literature has suggested that in settings with staggered treatment, the two-way fixed effects (TWFE) model may yield biased estimates. This bias arises from invalid comparisons, particularly when early-treated units are used as controls for later-treated units (Baker et al. 2022). To address this concern, we implement two alternative estimators that are robust to treatment effect heterogeneity: the two-stage DID estimator proposed by Gardner (2022) and the group-time average treatment effect estimator proposed by Callaway and Sant'Anna (2021). The results, as reported in Tables B1 and B2 in online Appendix B, are qualitatively consistent with our baseline findings. Specifically, the treatment effects of PE on care value are negative, marked by a reduction in operating costs and lower quality outcomes.

Lastly, one potential concern is that PE-owned hospitals may differ systematically in unobserved characteristics, such as managerial quality, market reputation, or latent financial health, which could bias our estimates. To address this concern, we implement a two-step Heckman selection model (Heckman 1979) to explicitly account for potential selection bias in PE acquisition decisions. In the first stage, we estimated a probit model of PE acquisition using all exogenous variables from the main specification, along with an instrumental variable that measures the number of prior-year successful IPOs in the focal hospital's Core-Based Statistical Area (CBSA), denoted as *PE_success*. This variable reflects local private capital activity and investment capacity that may influence PE acquisition likelihood but is plausibly exogenous to individual hospital performance. We then compute the inverse Mills ratio (IMR) from the first stage and include it as a correction term in the second-stage panel regression with hospital and year fixed effects (Wooldridge 2010). The inclusion of the IMR allows us to adjust for unobserved factors influencing both PE selection and hospital performance, thereby mitigating potential endogeneity concerns. The second-stage results, as reported in Table A4, indicate that PE ownership is associated with an approximately 0.11 decline in healthcare value, consistent with our baseline DID estimates.

**6.2 Moderating Role of Health Information Sharing**

In Table 3, we report our 2SLS estimation results of equation (3) as discussed in section 5.4.2. The 2SLS estimates of *Post×Treatment×InfoShare* suggest that PE-owned hospitals with higher levels of health information sharing exhibit higher care value than those with lower levels of health information sharing (coefficient=0.293, p-value < 0.01). The last few rows of Table 3 report the marginal effects of PE at different levels of health information sharing. Specifically, as shown in column (1), in hospitals with low levels of health information sharing (i.e., at $25^{th}$ percentile of *InfoShare*), PE ownership is associated with a 0.094 (i.e., 30.0%) decrease in care value. In contrast, the marginal effect at high levels of health information sharing (i.e., at $75^{th}$ percentile of *InfoShare*) is positive and significant. The difference in marginal effects between hospitals with high and low levels of health information sharing is also statistically significant (coefficient=0.161; p-value < 0.01).

To explore how health information sharing facilitates PE-owned hospitals' cost reduction strategies as discussed in H2, we examine the effect of PE ownership as well as the moderating role of health information sharing on measures of operating cost, non-labor cost, and labor cost. As reported in columns (2) to (4) of Table 3, the estimates of *Post×Treatment×InfoShare* suggest that the relationship between PE ownership and reduction in operating costs (including labor and non-labor costs) is more pronounced in hospitals characterized by high information sharing. Consistent with H2, these results suggest health information sharing may enable PE-owned hospitals to achieve greater cost reduction, which in turn contributes to higher care value.

To provide empirical evidence on how health information sharing facilitates PE-owned hospitals' care quality improvement, we use a range of quality metrics as the outcome variable and report the results in columns (5) to (7) in Table 3. As shown in column (7), when we measure quality using readmission rate, there is no significant relationship between PE ownership and readmission rate among hospitals with low levels of health information sharing (i.e., at $25^{th}$ percentile of *InfoShare*). However, in hospitals with greater information sharing (i.e., at $75^{th}$ percentile of *InfoShare*), PE ownership is associated with a statistically significant 0.52 (i.e., 2.63%) reduction in readmission rate. These findings support H3 and suggest that enhanced data access and care coordination facilitated by health information sharing allow PE-owned hospitals to effectively reduce readmission rates, which in turn contributes to higher care value.

However, when we use patient experiential quality as the quality metric, we find no evidence of a significant relationship between PE ownership and patient experiential quality in hospitals with either high or low levels of health information sharing, as shown in column (5) of Table 3. The result is not surprising. First, patient experiential quality is largely affected by patient-provider communication and staff responsiveness when patients are in the hospital, and is unlikely to be affected by health information sharing between providers (Joos et al. 2006; Kazley et al. 2012). When there are staffing reductions and shifts toward lower-cost, less-experienced clinical personnel following PE ownership (Borsa et al. 2023), the resulting worsened experiential quality cannot be easily improved by health information sharing.

Similarly, we find no significant relationship between PE ownership and mortality rate in hospitals with high health information sharing, as observed in column (6) of Table 3. This finding is consistent with the literature, which generally shows a limited role of health information sharing in reducing mortality (Hersh et al. 2015; Chen et al. 2019). There are several plausible explanations for this result. First, mortality is influenced by a wide range of complex factors, including in-hospital care processes and personnel mix, the severity of acute illness, comorbidities, and social determinants of health (Lantz et al., 1998). While information sharing may enhance communication between providers for better continuity of care, meaningful reduction in mortality requires systemic improvements in clinical pathways that may not be achieved solely through enhanced patient data sharing (Every et al. 2000; Shojania and Grimshaw 2005).

To further test the robustness of our 2SLS results, we employ an alternative IV to estimate the moderating effect of health information sharing. Specifically, we use the strength of broader inter-hospital IT infrastructure of neighboring hospitals in the same HRR, denoted as *Neighbor Inter-Hospital IT*, as an alternate IV. The results, as reported in Table A5 of the Appendix, are qualitatively consistent. Taken together, our results indicate that while PE ownership may negatively impact care value, these risks can be effectively alleviated in hospitals with IT-enabled health information sharing. Such information sharing can facilitate PE-owned hospitals' cost reduction goals; it can also support their efforts to improve readmission rates. Therefore, hospitals that prioritize robust health information sharing are better positioned to balance the financial pressures of PE ownership with the need to maintain high standards of patient care.

### 6.3 Mechanisms

#### 6.3.1 Health Information Sharing and Operating Cost Reduction

Our H2, along with the results in columns (2) to (4) of Table 3, suggests that health information sharing can facilitate reduction in operating costs following PE acquisition. We argue this is because higher levels of health information sharing could help PE-owned hospitals to implement more efficient workflows and reduce the amount of labor-intensive work, which lowers operating costs. To test this mechanism, we use workflow efficiency as the alternative outcome variable for model (3). Workflow efficiency is measured by the logged annual number of surgeries per unit of labor performed at a hospital. A higher value means hospital staff can complete more procedures per unit level of input, indicating greater workflow efficiency. Note that the annual number of surgeries per unit of labor could, in principle, be influenced by market demand factors. To account for patient demand variation, we include the log of the number of inpatient admissions as additional control variable.

Column (1) of Table 4 presents the 2SLS estimates of this model. The *Post×Treatment* coefficient is significantly negative (coefficient: -0.652; $p < 0.05$). In contrast, the coefficient of *Post×Treatment×InfoShare* is positive and statistically significant (coefficient: 0.983; $p < 0.05$). The estimated marginal effects of PE with different levels of health information sharing also suggest a positive relationship between PE ownership and workflow efficiency when the level of health information sharing is high, but the relationship becomes negative and marginally significant when the level of health information sharing is low. These findings provide preliminary support for our argument that health information sharing plays a critical role in enabling PE hospitals to enhance workflow efficiency, thereby reducing operating cost and improving care value.

### 6.3.2 Health Information Sharing and Care Quality Improvement

H3 discusses the moderating effect of health information sharing in facilitating PE-owned hospitals' efforts to improve care quality. A plausible mechanism is that by leveraging patient information shared by other providers, a PE-owned hospital can reduce adverse events and improve patient safety without excessive investment in proprietary information exchange infrastructure. This improves care quality and, in turn, care value. To test this mechanism, we use the standardized infection ratio (SIR) of healthcare associated infections (HAIs) as an alternate dependent variable for model (3). The SIR compares the observed number of HAIs to the predicted number, adjusting for risk factors known to be associated with infection incidence. A higher SIR indicates more infections than expected, reflecting poorer patient safety and more adverse events (Haque et al. 2020; Bao and Bardhan 2022). To measure HAIs, we use data from the CMS Hospital Compare database and include six types of infections: (1) central line-associated bloodstream infections in ICUs, (2) catheter-associated urinary tract infections in ICUs, (3) surgical site infections from colon surgery, (4) surgical site infections from abdominal hysterectomy, (5) methicillin-resistant Staphylococcus aureus (MRSA) blood laboratory-identified events, and (6) Clostridium difficile (C. diff) laboratory-identified events. We compute the average SIR across these six categories to construct our dependent variable, *Ratio of Adverse Events*.

Column (2) of Table 4 reports the 2SLS estimates of the model. The *Post*×*Treatment*×*InfoShare* coefficient is negative and statistically significant (coefficient: -1.157; $p < 0.05$). The last few rows of Table 4 report the marginal effects of PE ownership at different levels of *InfoShare*. Specifically, PE ownership is associated with a 0.276 increase in average SIR of HAIs at hospitals with low information sharing (i.e., $25^{th}$ percentile of *InfoShare*). This finding is consistent with the prior research (Kannan et al. 2023). However, the marginal effect of PE on hospitals with high levels of health information sharing (i.e., 75th percentile of *InfoShare*) is both negative and statistically significant. Specifically, in hospitals with high health information sharing, PE ownership is associated with a 0.361 reduction in the average SIR of HAIs. Overall, these results support our argument that health information sharing enables PE-owned hospitals to improve patient safety and reduce adverse events, which, in turn, improves care quality and care value.

**6.4 Heterogeneity Analyses**

Health information sharing occurs in two primary ways: (a) between hospitals, and (b) between hospitals and ambulatory care providers. Ambulatory providers, such as primary care physicians, outpatient specialists, and rehabilitation centers, are responsible for delivering outpatient care, including preventive services, diagnostics, minor procedures, and follow-up care (Martin-Misener et al. 2015). When hospitals efficiently share patient health information with ambulatory providers, both parties gain access to accurate and timely patient health information. This can enable them to treat patients effectively (Hesselink et al. 2014). For instance, prompt notification of a patient's discharge from the hospital allows ambulatory providers to coordinate follow-up appointments, tests, or consultations, which helps identify and manage potential issues early on (Jencks et al. 2009). Furthermore, timely access to hospital data enables ambulatory providers to use remote monitoring tools or conduct in-person assessments to detect early signs of deterioration in high-risk patients, allowing for preventive care before conditions worsen and require hospitalization (Jencks et al., 2009). Hence, effective health information sharing with ambulatory providers strengthens the complementarity between the focal hospital and ambulatory care providers, which may avoid unnecessary readmissions. In other words, we expect that sharing health information with ambulatory providers could help PE-owned hospitals to reduce operating costs and improve care quality, thereby improving care value.

On the other hand, sharing health information between hospitals serves a different function, often necessitated by patient transfers across care settings (Everson and Adler-Milstein 2020; Eftekhari et al. 2023). Hospital-to-hospital information sharing typically involves providing access to patient health data so that healthcare providers are aware of patients' prior hospitalization history and test results, and do not have to repeat such tests and procedures. It does not impact follow-up, care decisions, such as medication management or lifestyle changes, which are more commonly managed by ambulatory care providers. That is, although hospital-to-hospital information sharing facilitates reducing duplication and labor costs, it may have a limited impact on patient care after hospital discharge.

In summary, while both types of health information sharing play vital roles in facilitating PE-owned hospitals to improve care value, the mechanisms through which they shape care value may be different. We posit that the positive effect of health information sharing in moderating the impact of PE ownership on care quality is primarily driven by improved information exchange between hospitals and ambulatory providers. On the other hand, the positive role of health information sharing in moderating PE's impact on operating cost reduction may be driven by both information sharing with other hospitals and information sharing with ambulatory providers.

To provide empirical evidence on the heterogeneous effect among different types of health information sharing, we decompose information sharing into two categories: hospital-to-hospital information sharing (*InfoShare_hos*) and hospital-to-ambulatory provider information sharing (*InfoShare_amb*). We replace *InfoShare* in equation (3) with these variables and utilize 2SLS estimation.

Table 5 presents the results on the moderating role of hospital-to-hospital information sharing (*InfoShare_hos*). The coefficient of *Post*×*Treatment*×*InfoShare_hos* in column (1) indicates that more information sharing with other hospitals moderates the decline in care value following PE acquisition. Furthermore, the coefficient of *Post*×*Treatment*×*InfoShare_hos* is significant in columns (2) to (4), where cost is used as the dependent variable, but not in columns (5) to (7), where care quality is used as the dependent variable. This is consistent with our argument that hospital-to-hospital information sharing shapes PE's effect on care value mainly through its role in reducing operating cost.

In comparison, Table 6 reports the 2SLS estimates of the moderating impact of *InfoShare_Amb* on care value. Like *InfoShare_Hos*, the coefficient of *Post*×*Treatment*×*InfoShare_Amb* in column (1) shows that higher health information sharing with ambulatory providers helps mitigate the negative impact of PE ownership on care value. Further, *InfoShare_Amb* has a significant moderating effect on readmission rate reduction, as shown in column (7). It also enables PE's cost reduction strategies, especially in reducing labor cost, as shown in columns (2) and (4). This finding highlights that sharing information with ambulatory providers could play an important role in helping PE-owned hospitals to improve care quality and reduce cost, both of which contribute to higher care value.[11]

## 7. Robustness Checks

### 7.1. Alternate DEA inputs and outputs

The baseline DEA model includes capital-related expenses associated with the acquisition, maintenance, and use of property, plant, and equipment dedicated to patient care. To ensure the robustness of our findings with respect to input selection, we also estimate an alternative DEA model using the number of staffed beds as a proxy for capital input (Kohl et al. 2019; Bardhan et al. 2023). Specifically, this specification includes operating expenses, the number of staffed beds, and labor (including physicians, nurses, and other clinicians) as inputs. The results based on this alternate model are reported in Table A6 in the Appendix. It shows consistent results with respect to the impact of PE ownership on care value, as well as on operating costs, labor, and quality-related outcomes. Moreover, the moderating effect of health information sharing remains consistent under this alternative specification, as reported in Table A7.

### 7.2. Alternative Explanation: Potential Changes in Patient Volume

A potential concern is that the observed reduction in operating costs and readmission rates following PE acquisition in hospitals may be confounded by changes in patient volume. Specifically, PE-owned hospitals may be admitting fewer patients, thereby improving outcomes. To assess this possibility, we examine changes in patient flow by using the number of inpatient visits and inpatient days as the dependent variable for models (2) and (3). Table A8 first reports the estimation results based on model (2), followed by the 2SLS results based on model (3) in columns (3) and (4). The coefficient estimate for *Post×Treatment* is statistically insignificant across using both inpatient visits and total inpatient days as the dependent variable. This suggests that PE ownership does not seem to influence the overall patient flow or inpatient resources expended. It is consistent with prior research that shows hospital choice is predominantly influenced by geographic proximity and insurance coverage, rather than ownership status (Victoor et al. 2012; Smith et al. 2018). The estimate of *Post×Treatment×InfoShare* in columns (3) and (4) also indicates that the moderating effect of health information sharing is insignificant. These findings indicate that observed improvements in hospital efficiency and quality are unlikely to be driven by reduction in patient visits or inpatient utilization.

[11] For robustness, we also replicate all analyses using OLS estimates (i.e. without the use of instrumental variable for health information sharing). The results are qualitatively similar and available upon request.

In sum, we conduct a series of analyses to evaluate the robustness of our findings. Table A9 in the Appendix provides a detailed summary of the potential concerns addressed by these robustness checks, along with the corresponding results.

## 8. Discussion

We study the impact of PE ownership on the value of care delivered by US hospitals using a multi-input, multi-output measure of care value. Further, we focus on the role of health information sharing in moderating such a relationship. Our findings make several contributions to the literature. First, our research contributes to the literature on the impact of PE in healthcare by shifting the focus from isolated outcomes to a holistic evaluation metric, i.e., care value - a concept that emphasizes the delivery of high-quality and cost-effective care. This approach unites stakeholders' interests and is especially relevant given the current shift toward VBC models, where the emphasis is on delivering efficient and effective care rather than service volume (Bardhan et al. 2023; Porter 2010). Our findings indicate that although restructuring efforts following PE acquisition may yield short-term cost savings, they often come at the expense of reduced care quality, leading to lower healthcare value.

Second, we add to the literature on health information sharing by highlighting the unique role of IT-enabled health information sharing in mitigating the quality-cost tradeoffs faced by PE-owned hospitals. We show that PE-owned hospitals with strong health information sharing capabilities could reduce operating costs and improve care quality, both of which lead to higher care value. We also demonstrate the heterogeneous effects of hospital-to-hospital and hospital-to-ambulatory provider information sharing on care value following PE acquisition. While previous studies have examined the role of information sharing in improving healthcare efficiency and quality (Bao and Bardhan 2022;

Janakiraman et al. 2023), how different types of information sharing can moderate PE's impact on acquired hospitals remains an underexplored area of research. We address this gap by differentiating the roles of hospital-to-hospital and hospital-to-ambulatory provider information sharing in the context of PE ownership. Specifically, both types of health information sharing facilitate cost reduction by reducing redundant tasks and streamlining clinical workflows. However, it is the exchange of information between hospitals and ambulatory providers that has the most significant impact on improving care quality (i.e., reducing readmission rates). We highlight the important complementary role of ambulatory providers in managing follow-up care, coordinating post-discharge interventions, and preventing complications, which in turn, can reduce hospital readmission rates.

Third, our research is related to the broader literature that studies how structural changes affect hospital operations and patient outcomes. Prior studies have analyzed the impact of regulatory mandates (such as hospital price transparency rule) and market coordination mechanisms (such as referral and inter-organizational care coordination) on patient choice and care patterns (Christensen et al. 2020; Eftekhari et al. 2023). Our study adds to this discussion by highlighting how PE ownership, as a unique type of organizational structural change, interacts with the information-sharing mechanism to shape the efficiency and quality dimensions of hospital performance.

Our research also has important implications for various stakeholders. For healthcare policymakers, our research highlights the importance of a holistic approach when evaluating the impact of PE ownership in healthcare. We emphasize the need for regulatory oversight and policy interventions to ensure that PE ownership is aligned with the interests of a broad set of stakeholders. For hospital administrators, our findings suggest the importance of health information sharing capabilities in mitigating the potential negative effect of PE ownership. Administrators in PE-owned hospitals may benefit from proactively strengthening their information sharing capabilities to support operational efficiency and clinical coordination.

For PE firms, our research suggests that investing in hospitals with robust health information sharing infrastructure not only enables cost reduction but also enhances care quality improvement. This alignment of financial and clinical outcomes can enhance both short-term performance and long-term value creation. For patients, our research offers evidence that PE ownership does not inherently diminish care value, particularly when supported by strong health IT capabilities. This implies that receiving care at PE-owned hospitals with advanced information sharing infrastructure may lead to better care outcomes compared to those without such capabilities. At the same time, these insights highlight the importance of continued oversight to ensure that common PE restructuring strategies do not compromise patient-centered care.

## 9. Conclusion

What is the impact of PE ownership on care value and how can health IT mitigate the potential negative effects of PE ownership? Ours is among the first to answer these questions, focusing specifically on the role of IT-enabled, health information sharing. We observe that PE ownership of US hospitals yields lower care value, as measured by the degree to which a hospital utilizes its input resources efficiently in the production of healthcare services that enhance patient health outcomes. We also observe that health information sharing across healthcare providers can moderate this relationship. Specifically, PE-backed hospitals that engage in high levels of health information sharing with other healthcare providers can achieve the dual goals of cost efficiency and high-quality care. We demonstrate the robustness of our results using a range of methods, such as instrumental variable estimation, falsification tests, and alternative estimators. Additional evidence suggests that the reduction in readmission rates following PE acquisition is primarily driven by information sharing between hospitals and ambulatory providers, whereas the reduction in operating cost following PE acquisition is driven by both hospital-to-hospital and hospital-to-ambulatory health information sharing.

Our research highlights the importance of information sharing as a strategic capability in healthcare. Hospitals that invest in strong information sharing infrastructure are better positioned to improve care value, even in the face of ownership changes and financial pressure associated with PE ownership. Nevertheless, our study has several limitations. First, our analysis focuses exclusively on PE ownership in general and acute care hospitals. While hospitals represent a significant portion of PE activity in the healthcare sector, these investments extend far beyond hospital buyouts and acquisitions. PE firms are also increasingly active in other types of care delivery organizations, such as urgent care clinics, nursing homes, rehabilitation centers, specialty hospitals, and dental service organizations (Borsa et al. 2023). These settings represent unique operational challenges, regulatory environments, and patient demographics, which may influence the impact of PE in different ways. Future research can evaluate the effects of PE ownership across various care settings for a more comprehensive understanding across the healthcare continuum.

Second, effective information sharing is critical to enhance care coordination, reduce medical errors, and improve patient outcomes. Significant barriers remain in achieving a high level of information sharing across healthcare organizations, such as technology incompatibility, lack of common data standards, and concerns over data privacy and security (Li et al. 2023). Future research could explore other dimensions of information sharing, such as adoption of interoperability standards, which have the potential to drive high-quality care and mitigate the negative impacts of PE ownership.

Third, PE hospitals may differ in their strategic priorities, such as AI adoption, digital transformation, or service-line expansion. Future research can further investigate the heterogeneous effects on care value based on various characteristics of PE acquisitions. Meanwhile, to ensure greater focus in the scope of our study, we restrict our locus of attention on the impact of PE ownership on care value. An interesting direction for future research is to explore how PE ownership shapes IT adoption decisions in the acquired hospitals.

Last, given the observational nature of our data, the endogeneity issues cannot be fully addressed. Additional studies, particularly using alternative data sources or non-U.S. healthcare settings, are necessary to validate and extend our findings. Despite these limitations, our research represents an important step toward evaluating the impact of PE ownership on care value and proposing potential solutions to mitigate its potential negative impact. Our study can help safeguard the quality of care delivered, ultimately benefiting patients, providers, and investors alike. Our research can thereby serve as a catalyst to open new avenues for future research at the intersection of healthcare finance and information technology.

## Tables & Figures

**Table 1: Descriptive statistics of Matched Sample**

| | Period: 2008-2020 | | | Pre-treatment Period | | | | | |
|---|---|---|---|---|---|---|---|---|---|
| | Matched Sample | | | Control | | Treatment | | Balance Test | |
| Variables | Number of Observations | Mean | Standard Deviation | Number of Hospitals | Mean | Number of Hospitals | Mean | Mean Diff | p-value |
| *Panel A: DEA Variables* | | | | | | | | | |
| Operating Cost ($ in millions) | 5096 | 136.539 | 162.125 | 313 | 126.443 | 115 | 108.976 | 17.467 | 0.248 |
| Capital Cost ($ in millions) | 5096 | 8.684 | 10.822 | 313 | 8.119 | 115 | 7.618 | 0.500 | 0.596 |
| Number of Physicians | 5096 | 11.375 | 24.430 | 313 | 10.185 | 115 | 9.944 | 0.242 | 0.891 |
| Number of Nurses | 5096 | 267.627 | 277.943 | 313 | 252.515 | 115 | 238.155 | 14.360 | 0.593 |
| Number of Other clinicians | 5096 | 46.858 | 48.801 | 313 | 45.319 | 115 | 42.105 | 3.213 | 0.472 |
| Experiential Quality (in %) | 5096 | 73.047 | 4.991 | 313 | 72.468 | 115 | 72.238 | 0.230 | 0.635 |
| Mortality Rate (in %) | 5096 | 13.139 | 1.981 | 313 | 12.609 | 115 | 12.636 | -0.028 | 0.868 |
| Readmission Rate (in %) | 5096 | 19.794 | 1.979 | 313 | 20.339 | 115 | 20.309 | 0.030 | 0.871 |
| Care Value | 5096 | 0.311 | 0.199 | 313 | 0.303 | 115 | 0.308 | -0.006 | 0.749 |
| *Panel B: Control Variables* | | | | | | | | | |
| Intra-hospital IT | 5096 | 0.773 | 0.244 | 313 | 0.717 | 115 | 0.701 | 0.016 | 0.429 |
| Log(Bed Total) | 5096 | 4.797 | 0.762 | 313 | 4.755 | 115 | 4.848 | -0.093 | 0.263 |
| Log(Average Daily Census) | 5096 | 3.806 | 1.018 | 313 | 3.773 | 115 | 3.913 | -0.140 | 0.202 |
| Log(Inpatient Days Total) | 5096 | 9.973 | 1.040 | 313 | 9.974 | 115 | 9.985 | -0.011 | 0.922 |
| Resident to Bed Ratio | 5096 | 0.042 | 0.137 | 313 | 0.045 | 115 | 0.024 | 0.021 | 0.140 |
| Wage Index | 5096 | 0.962 | 0.184 | 313 | 0.968 | 115 | 0.940 | 0.028 | 0.131 |
| Case Mix Index | 5096 | 1.370 | 0.264 | 313 | 1.313 | 115 | 1.360 | -0.047 | 0.075 |
| Outlier Adjustment Factor | 5096 | 0.025 | 0.038 | 313 | 0.029 | 115 | 0.022 | 0.007 | 0.074 |
| *Panel C: Other Variables* | | | | | | | | | |
| Inter-hospital IT | 5096 | 0.541 | 0.251 | 313 | 0.481 | 115 | 0.477 | 0.004 | 0.836 |
| Non-Labor Operating Cost ($ in millions) | 5096 | 84.975 | 97.189 | 313 | 76.991 | 115 | 69.191 | 7.800 | 0.381 |
| Labor | 5096 | 325.861 | 328.105 | 313 | 308.019 | 115 | 290.204 | 17.815 | 0.573 |

Note: The descriptive statistics are reported based on the matched panel. Variables involved in DEA measurement model are listed in the Panel A. Variables used as controls are listed in Panel B. Panel C lists other variables utilized in the paper. We performed one-to-three PSM based on mean value of 8 hospital characteristics including number of physicians and nurses, experiential quality, readmission rate, mortality rate, measured healthcare value, and strength of intra-hospital IT infrastructure. The balance test is conducted by comparing the mean values of the eight variables for the years preceding the PE acquisition.

**Table 2: The Impact of PE Ownership on Care Value**

| | (1) | (2) | (3) | (4) | (5) | (6) | (7) |
|---|---|---|---|---|---|---|---|
| Dependent variable | Care Value | Operating Cost | Non-Labor Operating Cost | Labor | Experiential Quality | Mortality Rate | Readmission Rate |
| Post | 0.030** | 5.447** | 3.575** | 2.661 | 0.275 | -0.103 | 0.034 |
| | (0.012) | (2.415) | (1.679) | (7.664) | (0.185) | (0.103) | (0.085) |
| Post×Treatment | -0.034** | -19.995*** | -11.931*** | -26.714** | -0.788*** | 0.266* | 0.017 |
| | (0.014) | (5.216) | (3.855) | (11.571) | (0.290) | (0.154) | (0.118) |
| Observations | 5,096 | 5,096 | 5,096 | 5,096 | 5,096 | 5,096 | 5,096 |
| R-squared | 0.109 | 0.265 | 0.265 | 0.134 | 0.316 | 0.516 | 0.557 |
| Number of Hospitals | 428 | 428 | 428 | 428 | 428 | 428 | 428 |
| Controls | YES | YES | YES | YES | YES | YES | YES |
| Hospital FE | YES | YES | YES | YES | YES | YES | YES |
| Year FE | YES | YES | YES | YES | YES | YES | YES |

Notes: Labor is the combination of the number of FTE physicians, nurses and other clinicians. Robust standard errors clustered at hospital level in parentheses. *** p<0.01, ** p<0.05, * p<0.1.

**Table 3: 2SLS Estimation of Moderating Impact of Health Information Sharing**

| Dependent variable | (1) Care Value | (2) Operating Cost | (3) Non-Labor Operating Cost | (4) Labor | (5) Experiential Quality | (6) Mortality Rate | (7) Readmission Rate |
|---|---|---|---|---|---|---|---|
| Post | 0.127*** | -13.428 | -9.588 | -52.490 | 0.514 | 0.449 | -0.276 |
| | (0.040) | (11.346) | (7.824) | (32.084) | (0.702) | (0.394) | (0.348) |
| Post×InfoShare | -0.154*** | 28.776 | 19.753 | 86.321* | -0.414 | -0.817 | 0.467 |
| | (0.055) | (18.555) | (12.606) | (50.118) | (1.063) | (0.576) | (0.529) |
| Post×Treatment | -0.226*** | 41.377 | 35.636 | 108.916* | -0.078 | -0.589 | 1.160* |
| | (0.078) | (31.502) | (25.577) | (60.433) | (1.534) | (0.915) | (0.659) |
| Post×Treatment×InfoShare | 0.293*** | -92.835** | -71.551** | -205.901** | -0.919 | 1.278 | -1.680* |
| | (0.113) | (44.825) | (35.178) | (94.295) | (2.273) | (1.342) | (0.930) |
| Observations | 4,692 | 4,692 | 4,692 | 4,692 | 4,692 | 4,692 | 4,692 |
| R-squared | 0.114 | 0.265 | 0.260 | 0.136 | 0.320 | 0.510 | 0.566 |
| Number of Hospitals | 374 | 374 | 374 | 374 | 374 | 374 | 374 |
| Controls | YES | YES | YES | YES | YES | YES | YES |
| Hospital FE | YES | YES | YES | YES | YES | YES | YES |
| Year FE | YES | YES | YES | YES | YES | YES | YES |
| Marginal Effect of PE | -0.094*** | -0.398 | 3.438 | 16.261 | -0.492 | -0.014 | 0.404 |
| at p25 InfoShare | (0.030) | (12.207) | (10.217) | (20.471) | (0.566) | (0.337) | (0.258) |
| Marginal Effect of PE | 0.067* | -51.457*** | -35.914*** | -96.985*** | -0.997 | 0.689 | -0.520* |
| at p75 InfoShare | (0.038) | (14.944) | (10.634) | (36.848) | (0.824) | (0.469) | (0.304) |
| Difference Between | 0.161*** | -51.059** | -39.353** | -113.246** | -0.506 | 0.703 | -0.924* |
| p75 and p25 of InfoShare | (0.062) | (24.654) | (19.348) | (51.862) | (1.250) | (0.738) | (0.511) |

Note: Labor is the combination of the number of FTE physicians, nurses and other clinicians. IV is the average level of health information sharing at the neighboring hospitals in the same HRR. Robust standard errors clustered at hospital level in parentheses. *** p<0.01, ** p<0.05, * p<0.1.

**Table 4: Mechanism Analyses**

| Dependent Variable | (1) Workflow Efficiency | (2) Number of Adverse Events |
|---|---|---|
| Post | 0.189 | -0.148 |
| | (0.252) | (0.162) |
| Post×InfoShare | -0.290 | 0.173 |
| | (0.412) | (0.228) |
| Post×Treatment | -0.652** | 0.796** |
| | (0.323) | (0.348) |
| Post×Treatment×InfoShare | 0.983** | -1.157** |
| | (0.474) | (0.492) |
| Observations | 4,692 | 4,602 |
| R-squared | 0.039 | 0.009 |
| Number of Hospitals | 374 | 365 |
| Controls | YES | YES |
| Hospital FE | YES | YES |
| Year FE | YES | YES |
| Marginal Effect of PE | -0.197* | 0.276** |
| at p25 InfoShare | (0.116) | (0.135) |
| Marginal Effect of PE | 0.307* | -0.361** |
| at p75 InfoShare | (0.160) | (0.161) |
| Difference Between | 0.504** | -0.636** |
| p75 and p25 of InfoShare | (0.243) | (0.271) |

Note: Robust standard errors clustered at hospital level in parentheses. *** p<0.01, ** p<0.05, * p<0.1. Besides using the controls specified in Panel B in Table 1, we also include the log number of inpatient admissions as an additional control variable for market demand in column (1).

**Table 5: 2SLS Estimation of Moderating Impact of Health Information Sharing with Hospitals**

| Dependent variable | (1) Care Value | (2) Operating Cost | (3) Non-Labor Operating Cost | (4) Labor | (5) Experiential Quality | (6) Mortality Rate | (7) Readmission Rate |
|---|---|---|---|---|---|---|---|
| Post | 0.114*** | -14.796 | -10.160 | -38.666 | 0.955 | 0.505 | -0.223 |
| | (0.036) | (9.940) | (6.625) | (31.009) | (0.650) | (0.351) | (0.291) |
| Post×InfoShare_hos | -0.146*** | 33.986* | 22.463* | 71.289 | -1.224 | -0.995* | 0.447 |
| | (0.054) | (18.647) | (12.221) | (52.700) | (1.090) | (0.562) | (0.487) |
| Post×Treatment | -0.208*** | 44.540 | 39.175 | 81.688 | -0.296 | -0.536 | 0.826 |
| | (0.065) | (30.559) | (24.728) | (54.943) | (1.442) | (0.811) | (0.596) |
| Post×Treatment×InfoShare_hos | 0.289*** | -109.808** | -86.433** | -185.715* | -0.867 | 1.284 | -1.331 |
| | (0.104) | (49.116) | (38.338) | (96.097) | (2.365) | (1.337) | (0.938) |
| Observations | 4,618 | 4,618 | 4,618 | 4,618 | 4,618 | 4,618 | 4,618 |
| R-squared | 0.099 | 0.253 | 0.245 | 0.134 | 0.317 | 0.507 | 0.561 |
| Number of Hospitals | 368 | 368 | 368 | 368 | 368 | 368 | 368 |
| Controls | YES | YES | YES | YES | YES | YES | YES |
| Hospital FE | YES | YES | YES | YES | YES | YES | YES |
| Year FE | YES | YES | YES | YES | YES | YES | YES |
| Marginal Effect of PE | -0.063*** | -10.364 | -4.042 | -11.169 | -0.730* | 0.106 | 0.161 |
| at p25 InfoShare_hos | (0.018) | (8.094) | (6.712) | (13.294) | (0.385) | (0.205) | (0.166) |
| Marginal Effect of PE | 0.081* | -65.269*** | -47.258*** | -104.027** | -1.163 | 0.747 | -0.504 |
| at p75 InfoShare_hos | (0.043) | (20.084) | (14.612) | (44.182) | (1.007) | (0.566) | (0.374) |
| Difference Between | 0.145*** | -54.904** | -43.217** | -92.857* | -0.434 | 0.642 | -0.665 |
| p75 and p25 of InfoShare_hos | (0.052) | (24.558) | (19.169) | (48.049) | (1.183) | (0.669) | (0.469) |

Note: Labor includes the number of FTE physicians, nurses and other clinicians. IV is the average level of health information sharing at the neighboring hospitals in the same HRR. Robust standard errors clustered at hospital level in parentheses. *** p<0.01, ** p<0.05, * p<0.1.

**Table 6: 2SLS Estimation of Moderating Impact of Health Information Sharing with Ambulatory Providers**

| Dependent variable | (1) Care Value | (2) Operating Cost | (3) Non-Labor Operating Cost | (4) Labor | (5) Experiential Quality | (6) Mortality Rate | (7) Readmission Rate |
|---|---|---|---|---|---|---|---|
| Post | 0.154*** | -8.430 | -6.927 | -68.552 | -0.062 | 0.366 | -0.396 |
| | (0.055) | (16.722) | (11.879) | (43.696) | (0.981) | (0.545) | (0.479) |
| Post×InfoShare_amb | -0.182** | 19.797 | 14.837 | 103.425* | 0.491 | -0.625 | 0.616 |
| | (0.073) | (24.086) | (17.009) | (62.680) | (1.392) | (0.760) | (0.677) |
| Post×Treatment | -0.236*** | 24.947 | 21.290 | 118.408* | 0.170 | -0.412 | 1.315* |
| | (0.085) | (29.344) | (23.141) | (62.785) | (1.554) | (0.914) | (0.681) |
| Post×Treatment×InfoShare_amb | 0.281** | -62.450* | -45.850 | -201.502** | -1.200 | 0.914 | -1.749** |
| | (0.112) | (37.811) | (28.915) | (88.343) | (2.095) | (1.218) | (0.886) |
| Observations | 4,675 | 4,675 | 4,675 | 4,675 | 4,675 | 4,675 | 4,675 |
| R-squared | 0.106 | 0.265 | 0.262 | 0.124 | 0.319 | 0.508 | 0.567 |
| Number of Hospitals | 372 | 372 | 372 | 372 | 372 | 372 | 372 |
| Controls | YES | YES | YES | YES | YES | YES | YES |
| Hospital FE | YES | YES | YES | YES | YES | YES | YES |
| Year FE | YES | YES | YES | YES | YES | YES | YES |
| Marginal Effect of PE | -0.095*** | -6.278 | -1.635 | 17.657 | -0.430 | 0.045 | 0.440* |
| at p25 InfoShare_hos | (0.032) | (11.351) | (9.196) | (21.203) | (0.562) | (0.332) | (0.257) |
| Marginal Effect of PE | 0.045 | -37.504*** | -24.560*** | -83.094*** | -1.030 | 0.503 | -0.434* |
| at p75 InfoShare_hos | (0.031) | (10.556) | (7.186) | (29.317) | (0.642) | (0.356) | (0.245) |
| Difference Between | 0.141** | -31.225* | -22.925 | -100.751** | -0.600 | 0.457 | -0.875** |
| p75 and p25 of InfoShare_amb | (0.056) | (18.905) | (14.457) | (44.172) | (1.047) | (0.609) | (0.443) |

Note: Labor includes the number of FTE physicians, nurses and other clinicians. Ambulatory providers include physician offices, ambulatory surgery centers, outpatient clinics, etc. IV is the average health information sharing at the neighboring hospitals in the same HRR. Robust standard errors clustered at hospital level in parentheses. *** p<0.01, ** p<0.05, * p<0.1.

# Online Appendix

**Table OA1: Survey Questions in AHA IT Supplement Data**

| | | |
|---|---|---|
| **Section 1: Intra-hospital IT Capabilities** | | |
| ***Does your hospital currently have a computerized system which allows for:*** | | |
| **1. Electronic Clinical Documentation** | **2. Results viewing** | **3. Computerized Provider Order Entry** |
| a. Physician notes | a. Laboratory tests | a. Laboratory tests |
| b. Nursing notes | b. Radiology tests | b. Radiology tests |
| c. Problem lists | c. Radiology images | c. Medications |
| d. Medication lists | d. Diagnostic test results (e.g. EKG report, Echo report) | d. Consultation requests |
| e. Discharge summaries | e. Diagnostic test images (e.g. EKG tracing) | e. Nursing orders |
| f. Advanced directives (e.g. DNR) | f. Consultant reports | |
| | | |
| **4. Decision support** | **5. Other functionalities** | |
| a. Clinical guidelines (e.g. Beta blockers post-MI, ASA in CAD) | a. Bar coding or Radio Frequency (RFID) for supply chain management | |
| b. Clinical reminders (e.g. pneumovax) | b. Telehealth | |
| c. Drug allergy alerts | c. Remote patient monitoring | |
| d. Drug-drug interaction alerts | | |
| e. Drug-Lab interaction alerts | | |
| f. Drug dosing support (e.g. renal dose guidance) | | |
| | | |
| **Section 2: Health Information Sharing with Healthcare Providers** | | |
| ***Which of the following patient data does your hospital electronically exchange/share with one or more of the provider types listed below?*** | | |
| **1. Patient demographics** | **2. Laboratory results** | **3. Medication history** |
| a. With Hospitals inside of your System | a. With Hospitals inside of your System | a. With Hospitals inside of your System |
| b. With Hospitals Outside of your system | b. With Hospitals Outside of your system | b. With Hospitals Outside of your system |
| c. With Ambulatory Providers inside of your system | c. With Ambulatory Providers inside of your system | c. With Ambulatory Providers inside of your system |
| d. With Ambulatory Providers outside of your system | d. With Ambulatory Providers outside of your system | d. With Ambulatory Providers outside of your system |
| | | |
| **4. Radiology reports** | **5. Clinical / Summary care record in any format** | |
| a. With Hospitals inside of your System | a. With Hospitals inside of your System | |
| b. With Hospitals Outside of your system | b. With Hospitals Outside of your system | |
| c. With Ambulatory Providers inside of your system | c. With Ambulatory Providers inside of your system | |
| d. With Ambulatory Providers outside of your system | d. With Ambulatory Providers outside of your system | |
| | | |
| **Section 3: Health Information Sharing with Patients** | | |
| ***Are patients treated in your hospital able to do the following:*** | | |
| a. View their health/medical information online | b. Download information from their health/medical record | c. Electronically transmit (send) transmission of care/referral summaries to a third party |
| d. Request an amendment to change/update their health/medical record | e. Request refills for prescriptions online | f. Schedule appointments online |
| g. Pay bills online | h. Submit patient-generated data (e.g. blood, glucose, weight) | i. Secure messaging with providers |

## Appendix A: Additional Empirical Results

**Figure A1: Results of Placebo Test**

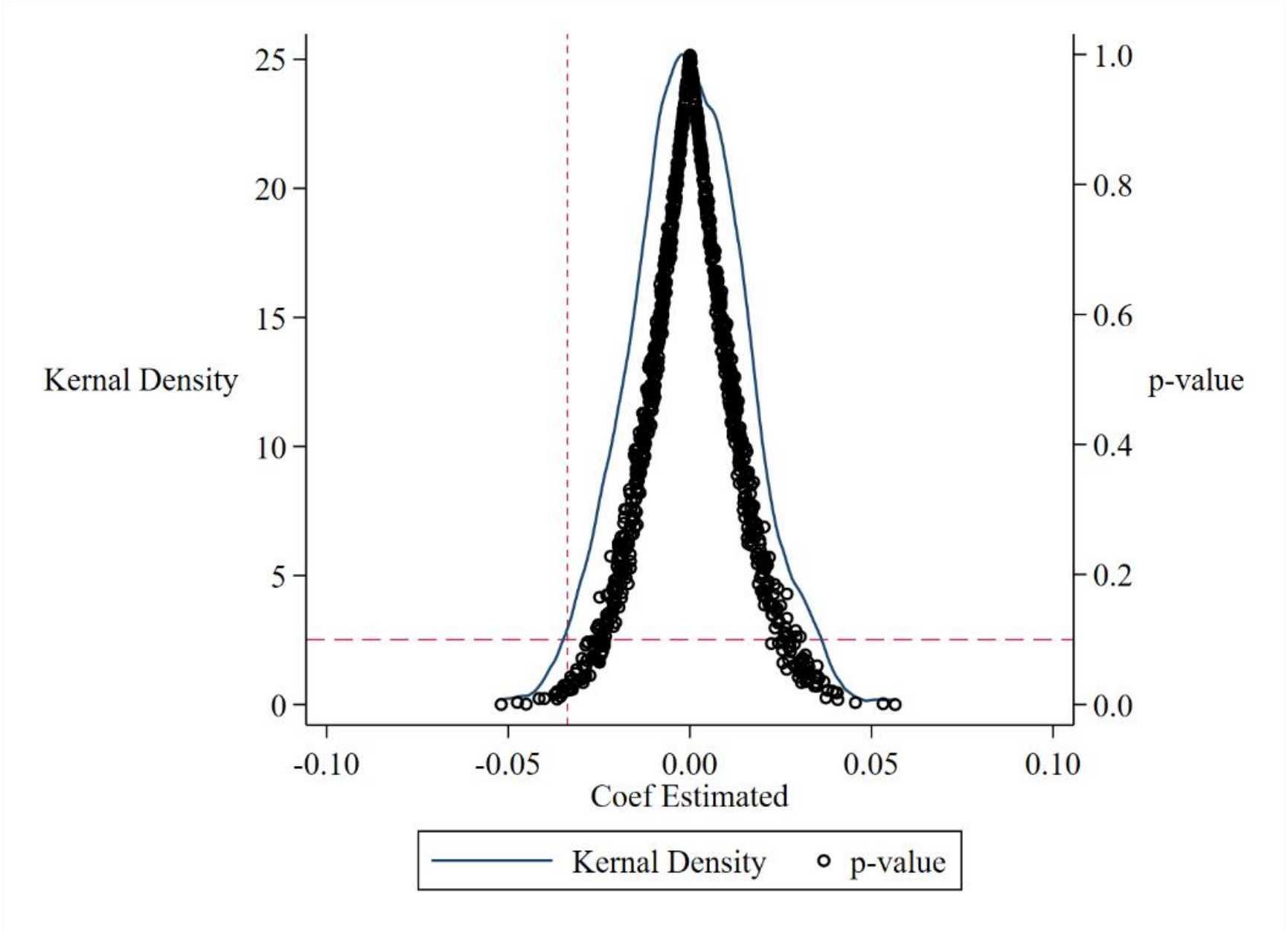

**Table A1: Exogeneity Test Results for IV**

| VARIABLES | (1)<br>Care Value | (2)<br>Log ( Inpatient Visit) | (3)<br>Log ( Inpatient Days) | (4)<br>Total Physician FTE | (5)<br>Total Nurse FTE | (6)<br>Total Other Clinician FTE | (7)<br>Labor |
|---|---|---|---|---|---|---|---|
| Neighbor InfoShare | -0.013 | -0.047 | -0.071 | -3.083 | -23.614 | -4.252 | -30.949 |
| | (0.026) | (0.043) | (0.047) | (3.586) | (17.220) | (3.770) | (20.502) |
| Observations | 5,082 | 5,082 | 5,082 | 5,082 | 5,082 | 5,082 | 5,082 |
| R-squared | 0.105 | 0.451 | 0.463 | 0.015 | 0.139 | 0.033 | 0.131 |
| Number of Hospitals | 426 | 426 | 426 | 426 | 426 | 426 | 426 |
| Controls | YES | YES | YES | YES | YES | YES | YES |
| Hospital FE | YES | YES | YES | YES | YES | YES | YES |
| Year FE | YES | YES | YES | YES | YES | YES | YES |

Note: In column (2) and (3), the log number of total inpatient days is excluded from the list of controls variables to avoid collinearity issue. Robust standard errors clustered at hospital level in parentheses. *** p<0.01, ** p<0.05, * p<0.1.

**Table A2: Double ML Estimation of PE Impact on Care Value**

| Dependent variable | (1)<br>Care Value |
|---|---|
| Post×Treatment | -0.026*** |
| | (0.009) |
| Observations | 5,092 |

Notes: Robust standard errors are given in parentheses. *** p<0.01, ** p<0.05, * p<0.1.

**Table A3: Regression Results for Care Value: Leads/Lags Model**

| Dependent Variable | Care Value | |
|---|---|---|
| | Coef. | Std. Error |
| Pre(-4)×Treatment | 0.010 | (0.014) |
| Pre(-3)×Treatment | -0.011 | (0.018) |
| Pre(-2)×Treatment | -0.021 | (0.018) |
| Pre(-1)×Treatment | -0.013 | (0.019) |
| Post(0)×Treatment | -0.033 | (0.022) |
| Post(1)×Treatment | -0.040** | (0.020) |
| Post(2)×Treatment | -0.067*** | (0.026) |
| Post(3)×Treatment | -0.023 | (0.051) |
| Post(4+)×Treatment | -0.110** | (0.046) |
| Observations | 5,096 | |
| Number of Hospitals | 428 | |
| R-squared | 0.115 | |
| Controls | YES | |
| Hospital FE | YES | |
| Year FE | YES | |

Note: Term "Pre(-*i*)×Treatment" indicates *i* years prior to the start year for the treatment group. The term "Post(*i*)×Treatment" indicates *i* years after the treatment start year for the treatment group. "Post(0)×Treatment" indicates the start year for the treatment group. "Pre(-5)×Treatment" is omitted baseline period. We do not report coefficients of "Pre(-*i*)" and "Post(*i*)" due to space limitations; complete regression results are available upon request. Robust standard errors clustered at hospital level in parentheses. *** p<0.01, ** p<0.05, * p<0.1.

**Table A4: Effect on PE Ownership on Healthcare Value using Heckman Selection**

| | (1) |
|---|---|
| Dependent Variable | Care Value |
| Post | 0.024* |
| | (0.012) |
| Post×Treatment | -0.109** |
| | (0.053) |
| IMR | 0.051* |
| | (0.030) |
| Observations | 4,287 |
| R-squared | 0.097 |
| Number of Hospitals | 422 |
| Controls | YES |
| Hospital FE | YES |
| Year FE | YES |

Robust standard errors clustered at hospital level in parentheses. *** p<0.01, ** p<0.05, * p<0.1.

**Table A5: 2SLS Estimation of Moderating Impact of Health Information Sharing with Alternative IV**

| Dependent variable | (1)<br>Care Value | (2)<br>Operating Cost | (3)<br>Non-Labor Operating Cost | (4)<br>Labor | (5)<br>Experiential Quality | (6)<br>Mortality Rate | (7)<br>Readmission Rate |
|---|---|---|---|---|---|---|---|
| Post | 0.146*** | -13.709 | -10.479 | -38.061 | 0.768 | 0.413 | -0.379 |
| | (0.040) | (12.534) | (8.960) | (34.635) | (0.713) | (0.416) | (0.375) |
| Post×InfoShare | -0.184*** | 29.516 | 21.449 | 62.434 | -0.808 | -0.763 | 0.640 |
| | (0.055) | (19.863) | (14.127) | (52.192) | (1.079) | (0.608) | (0.569) |
| Post×Treatment | -0.276*** | 32.846 | 28.373 | 117.694* | -0.702 | -0.453 | 1.053* |
| | (0.080) | (28.671) | (22.185) | (65.245) | (1.530) | (0.936) | (0.627) |
| Post×Treatment×InfoShare | 0.369*** | -80.096** | -60.720** | -218.721** | 0.018 | 1.074 | -1.522* |
| | (0.115) | (40.804) | (30.497) | (99.369) | (2.275) | (1.362) | (0.881) |
| Observations | 4,692 | 4,692 | 4,692 | 4,692 | 4,692 | 4,692 | 4,692 |
| R-squared | 0.107 | 0.269 | 0.267 | 0.133 | 0.320 | 0.511 | 0.566 |
| Number of Hospitals | 374 | 374 | 374 | 374 | 374 | 374 | 374 |
| Controls | YES | YES | YES | YES | YES | YES | YES |
| Hospital FE | YES | YES | YES | YES | YES | YES | YES |
| Year FE | YES | YES | YES | YES | YES | YES | YES |
| Marginal Effect of PE | -0.110*** | -3.198 | 1.049 | 19.269 | -0.694 | 0.031 | 0.368 |
| at p25 InfoShare | (0.030) | (11.233) | (8.999) | (22.858) | (0.562) | (0.348) | (0.248) |
| Marginal Effect of PE | 0.092** | -47.251*** | -32.347*** | -101.027*** | -0.684 | 0.621 | -0.469 |
| at p75 InfoShare | (0.039) | (13.836) | (9.482) | (37.268) | (0.829) | (0.467) | (0.288) |
| Difference Between | 0.203*** | -44.053** | -33.396** | -120.297** | 0.010 | 0.591 | -0.837* |
| p75 and p25 Marginal Effect | (0.063) | (22.442) | (16.773) | (54.653) | (1.251) | (0.749) | (0.485) |

Note: Labor is the combination of the number of FTE physicians, nurses and other clinicians. IV is the average level of *Inter-Hospital IT* at the neighboring hospitals in the same HRR. Robust standard errors clustered at hospital level in parentheses. *** p<0.01, ** p<0.05, * p<0.1.

**Table A6: Impact of PE on Care Value using Alternate DEA inputs and outputs**

| Dependent variable | (1)<br>Care Value | (2)<br>Operating Cost | (3)<br>Non-Labor Operating Cost | (4)<br>Labor | (5)<br>Experiential Quality | (6)<br>Mortality Rate | (7)<br>Readmission Rate |
|---|---|---|---|---|---|---|---|
| Post | 0.020* | 2.621 | 2.201 | 11.161 | 0.421** | -0.221** | -0.013 |
| | (0.010) | (2.987) | (2.227) | (7.376) | (0.199) | (0.102) | (0.086) |
| Post×Treatment | -0.028** | -14.543*** | -9.234** | -20.244* | -0.737** | 0.188 | -0.059 |
| | (0.013) | (5.117) | (4.046) | (10.414) | (0.294) | (0.151) | (0.118) |
| Observations | 5,080 | 5,080 | 5,080 | 5,080 | 5,080 | 5,080 | 5,080 |
| R-squared | 0.212 | 0.228 | 0.054 | 0.214 | 0.024 | 0.134 | 0.050 |
| Number of Hospitals | 422 | 422 | 422 | 422 | 422 | 422 | 422 |
| Controls | YES | YES | YES | YES | YES | YES | YES |
| Hospital FE | YES | YES | YES | YES | YES | YES | YES |
| Year FE | YES | YES | YES | YES | YES | YES | YES |

Notes: Labor is the combination of the number of FTE physicians, nurses and other clinicians. Robust standard errors clustered at hospital level in parentheses. *** p<0.01, ** p<0.05, * p<0.1.

**Table A7: 2SLS Estimation of Moderating Impact of Health Information Sharing using Alternate DEA Inputs and Outputs**

| Dependent variable | (1) Care Value | (2) Operating Cost | (3) Non-Labor Operating Cost | (4) Labor | (5) Experiential Quality | (6) Mortality Rate | (7) Readmission Rate |
|---|---|---|---|---|---|---|---|
| Post | 0.061 | -1.292 | -3.919 | -5.961 | 0.901 | 0.527 | -0.471 |
| | (0.040) | (17.645) | (12.835) | (35.485) | (0.989) | (0.459) | (0.408) |
| Post×InfoShare | -0.067 | 4.555 | 8.300 | 26.125 | -0.839 | -1.148* | 0.707 |
| | (0.059) | (27.495) | (19.847) | (57.181) | (1.449) | (0.677) | (0.601) |
| Post×Treatment | -0.179** | 37.240 | 35.476 | 62.179 | 0.084 | -0.865 | 1.299* |
| | (0.079) | (34.624) | (28.045) | (55.318) | (1.715) | (0.933) | (0.676) |
| Post×Treatment×InfoShare | 0.229** | -77.396 | -66.676* | -125.112 | -1.109 | 1.604 | -2.010** |
| | (0.114) | (48.672) | (38.684) | (86.077) | (2.530) | (1.364) | (0.962) |
| Observations | 4,757 | 4,757 | 4,757 | 4,757 | 4,757 | 4,757 | 4,757 |
| R-squared | 0.224 | 0.214 | 0.197 | 0.141 | 0.272 | 0.529 | 0.555 |
| Number of Hospitals | 381 | 381 | 381 | 381 | 381 | 381 | 381 |
| Controls | YES | YES | YES | YES | YES | YES | YES |
| Hospital FE | YES | YES | YES | YES | YES | YES | YES |
| Year FE | YES | YES | YES | YES | YES | YES | YES |
| Marginal Effect of PE | -0.065*** | -1.458 | 2.139 | -0.376 | -0.470 | -0.063 | 0.294 |
| at p25 InfoShare | (0.024) | (11.271) | (9.357) | (15.343) | (0.520) | (0.284) | (0.219) |
| Marginal Effect of PE | 0.050 | -40.156*** | -31.199*** | -62.932* | -1.025 | 0.739 | -0.711** |
| at p75 InfoShare | (0.038) | (15.482) | (11.698) | (33.398) | (0.895) | (0.470) | (0.318) |
| Difference Between | 0.115** | -38.698 | -33.338* | -62.556 | -0.554 | 0.802 | -1.005** |
| p75 and p25 Marginal Effect | (0.057) | (24.336) | (19.342) | (43.039) | (1.265) | (0.682) | (0.481) |

Note: Labor is the combination of the number of FTE physicians, nurses and other clinicians. IV is the average level of health information sharing at the neighboring hospitals in the same HRR. Robust standard errors clustered at hospital level in parentheses. *** p<0.01, ** p<0.05, * p<0.1.

**Table A8: Impact of PE Ownership and Information Sharing on Inpatient Volume and Inpatient Days**

| Dependent Variable | (1) Log(Inpatient Admissions) | (2) Log(Inpatient Days) | (3) Log(Inpatient Admissions) | (4) Log(Inpatient Days) |
|---|---|---|---|---|
| Post | 0.003 | 0.013 | 0.019 | 0.048 |
| | (0.015) | (0.019) | (0.058) | (0.063) |
| Post×InfoShare | | | -0.023 | -0.051 |
| | | | (0.085) | (0.095) |
| Post×Treatment | 0.016 | 0.020 | 0.144 | -0.013 |
| | (0.022) | (0.026) | (0.119) | (0.142) |
| Post×Treatment×InfoShare | | | -0.198 | 0.048 |
| | | | (0.169) | (0.207) |
| Observations | 5,096 | 5,096 | 4,692 | 4,692 |
| R-squared | 0.452 | 0.463 | 0.450 | 0.473 |
| Number of Hospitals | 428 | 428 | 374 | 374 |
| Controls | YES | YES | YES | YES |
| Hospital FE | YES | YES | YES | YES |
| Year FE | YES | YES | YES | YES |

Note: We use log transformation to reduce the skewness of the dependent variable. Columns (3) and (4) are 2SLS results using the average level of health information sharing at the neighboring hospitals in the same HRR. In all columns, the log number of total inpatient days is excluded from the list of control variables to avoid collinearity. Robust standard errors clustered at hospital level in parentheses. *** p<0.01, ** p<0.05, * p<0.1.

**Table A9: Summary of Robustness Checks**

| Concerns | Tests and Results of additional analyses | Location |
|---|---|---|
| The exogeneity assumption of our IV may not be valid. | Test: Exogeneity test using care value, patient flow, and staff level.<br>Findings: IV does not have a direct effect on care value, patient flow, and staff level of the local hospital, suggesting that the exogeneity assumption of our IV is valid. | Table A1 |
| Potential endogeneity may arise from complex interactions or nonlinear relationships among covariates, which may not be fully captured by the parametric models (e.g., logistic regression) typically used in PSM. | Test: Double machine learning to more flexibly control for covariates that may jointly affect treatment selection and outcomes.<br>Findings: Result remain consistent with that reported in Table 2. | Table A2 |
| There may be substantial differences between the treated and control hospitals in the years prior to the treatment. | Test: Whether treated and control hospitals exhibit parallel trends.<br>Findings: Results support the validity of the parallel trend assumption. | Table A3 |
| Observed effects may be due to pre-existing trends or spurious correlations, rather than the actual treatment. | Test: Placebo tests are conducted by randomly assigning treated hospitals.<br>Findings: The DID estimator does not systematically generate spurious treatment effects. | Figure A2 |
| In settings with staggered treatment timing, the two-way fixed effects model may yield biased estimates | Test: Alternative estimators that are robust to treatment effect heterogeneity.<br>Findings: Results are similar to those reported in Table 2. | Table B1 & Table B2 |
| Unobserved variables may bias the estimate of PE's impact even though we employ methods such as PSM and DML, which mainly rely on observed characteristics. | Test: A two-step Heckman selection model<br>Findings: Results are consistent with those reported in Table 2. | Table A4 |
| The validity of the original IV strategy | Test: Alternate IV using the average level of Inter-hospital IT of neighboring hospitals in the same HRR.<br>Findings: Results remain qualitatively consistent with those reported in Table 3. | Table A5 |
| The selection of DEA input variables may drive our results | Test: Alternate DEA input variables using the number of staffed beds.<br>Findings: Results are similar to those reported in Tables 2 and 3. | Table A6 & Table A7 |
| Observed reductions in operating costs and readmission rates after PE acquisition may be confounded by changes in patient volume. | Test: Examine changes in patient flow before and after PE acquisition .<br>Findings: PE acquisition does not significantly alter overall patient flow. | Table A8 |

## Appendix B: Heterogeneity Robust Estimators

In staggered DID settings, the standard TWFE estimator can yield biased estimates, particularly when treatment effects are heterogeneous across units or vary over time (Baker et al. 2022). To address this issue, we implement two alternative estimation strategies that are robust to treatment effect heterogeneity. First, we apply the two-stage DID estimator introduced by Gardner (2022). This method mitigates the bias by residualizing the outcome variable in the first stage through a regression on unit and time fixed effects. In the second stage, the residualized outcome is regressed on the treatment indicator, isolating the treatment effect while accounting for fixed heterogeneity. We apply this estimator to our baseline model (i.e., equation (2)), and the results are presented in Table B1. They are qualitatively similar to the ones reported in our baseline table.

**Table B1: Heterogeneity Robust Estimation of Impact of PE on Care Value**

| Dependent variable | (1)<br>Care Value | (2)<br>Operating Cost | (3)<br>Non-Labor Operating Cost | (4)<br>Labor | (5)<br>Experiential Quality | (6)<br>Mortality Rate | (7)<br>Readmission Rate |
|---|---|---|---|---|---|---|---|
| Post×Treatment | -0.043** | -19.47*** | -11.03** | -19.23 | -1.044** | 0.272 | 0.011 |
| | (0.015) | (5.324) | (4.066) | (11.97) | (0.327) | (0.154) | (0.12) |
| Observations | 5,096 | 5,096 | 5,096 | 5,096 | 5,096 | 5,096 | 5,096 |
| Number of Hospitals | 428 | 428 | 428 | 428 | 428 | 428 | 428 |
| Controls | YES | YES | YES | YES | YES | YES | YES |
| Hospital FE | YES | YES | YES | YES | YES | YES | YES |
| Year FE | YES | YES | YES | YES | YES | YES | YES |

Notes: Labor is the combination of the number of FTE physicians, nurses and other clinicians. Robust standard errors clustered at hospital level in parentheses. *** p<0.01, ** p<0.05, * p<0.1.

Second, we implement the group-time average treatment effect estimator proposed by Callaway and Sant'Anna (2021). This method identifies valid control groups for each treated cohort and time period and estimates group-time-specific average treatment effects. These effects are aggregated to obtain the overall average treatment effect. We apply this estimator to the same baseline model, and the corresponding results are reported in Table B2. These results are consistent with our baseline findings.

**Table B2: Alternate Heterogeneity Robust Estimation of Impact of PE on Care Value**

| Dependent variable | (1)<br>Care Value | (2)<br>Operating Cost | (3)<br>Non-Labor Operating Cost | (4)<br>Labor | (5)<br>Experiential Quality | (6)<br>Mortality Rate | (7)<br>Readmission Rate |
|---|---|---|---|---|---|---|---|
| Post×Treatment | -0.060** | -17.24* | -11.34* | 3.320 | -0.691 | 0.442* | 0.140 |
| | (0.021) | (6.923) | (5.069) | (14.99) | (0.423) | (0.176) | (0.128) |
| Observations | 5,084 | 5,084 | 5,084 | 5,084 | 5,084 | 5,084 | 5,084 |
| Number of Hospitals | 425 | 425 | 425 | 425 | 425 | 425 | 425 |
| Controls | YES | YES | YES | YES | YES | YES | YES |
| Hospital FE | YES | YES | YES | YES | YES | YES | YES |
| Year FE | YES | YES | YES | YES | YES | YES | YES |

Notes: Labor is the combination of the number of FTE physicians, nurses and other clinicians. Robust standard errors clustered at hospital level in parentheses. *** p<0.01, ** p<0.05, * p<0.1.